\documentclass[conference]{IEEEtran}

\usepackage[T1]{fontenc}
\usepackage[utf8]{inputenc}
\usepackage{cite}
\usepackage{amsmath,amssymb}
\usepackage{graphicx}
\usepackage{booktabs}
\usepackage{tabularx}
\usepackage{multirow}
\usepackage{algorithm}
\usepackage{algpseudocode}
\usepackage{listings}
\usepackage{xcolor}
\usepackage{url}
\usepackage{hyperref}
\usepackage{balance}
\usepackage{microtype}
\usepackage{tikz}
\usetikzlibrary{shapes.geometric, arrows.meta, positioning, fit, backgrounds}

\lstdefinestyle{nginxstyle}{
  backgroundcolor=\color{gray!8},
  basicstyle=\ttfamily\scriptsize,
  breaklines=true,
  frame=single,
  rulecolor=\color{gray!40},
  captionpos=b,
  numbers=left,
  numberstyle=\tiny\color{gray},
  keywordstyle=\color{blue!70}\bfseries,
  commentstyle=\color{green!50!black},
  stringstyle=\color{orange!80!black},
}
\hypersetup{
  colorlinks=true,
  linkcolor=black,
  citecolor=black,
  urlcolor=blue!70!black
}

\title{Operationalising Post‑Quantum TLS: Automated Configuration Profiling and Hybrid PQC Deployment in Financial Infrastructure}

\author{
  \IEEEauthorblockN{%
    Harish Balaji\IEEEauthorrefmark{1},
    Aarav Varshney\IEEEauthorrefmark{2}\IEEEauthorrefmark{3},
    Prasanna Ravi\IEEEauthorrefmark{2}\IEEEauthorrefmark{3},
    Sripal Jain\IEEEauthorrefmark{4},
    Robin Foe\IEEEauthorrefmark{4},\\
    Jorden Seet\IEEEauthorrefmark{4},
    Huaxiong Wang\IEEEauthorrefmark{1},
    Kwok-Yan Lam\IEEEauthorrefmark{1},
    Anupam Chattopadhyay\IEEEauthorrefmark{2}\IEEEauthorrefmark{3}
  }

      \IEEEauthorblockA{%
    \IEEEauthorrefmark{1}%
    Digital Trust Centre, Nanyang Technological University, Singapore\\
    \texttt{\{harish.balaji, HXWang, kwokyan.lam\}@ntu.edu.sg}%
  }

   \IEEEauthorblockA{%
    \IEEEauthorrefmark{2}%
    College of Computing and Data Science, Nanyang Technological University, Singapore\\
    \texttt{\{aarav.varshney, prasanna.ravi, anupam\}@ntu.edu.sg}%
  }
  
  \IEEEauthorblockA{%
    \IEEEauthorrefmark{3}%
    PQStation, Singapore\\
    \texttt{\{aarav, prasanna\}@pqstation.com}%
  }

  \IEEEauthorblockA{%
    \IEEEauthorrefmark{4}%
    OCBC Bank, Singapore\\
    \texttt{\{sripaljain, jordenseet, robinfoe\}@ocbc.com}%
  }
  
}

\begin{document}

\maketitle


\begin{abstract}

Organisations are upgrading their cryptographic infrastructure to become quantum‑safe before large‑scale quantum computers materialise. Post‑quantum cryptography (PQC) standards now exist for key‑exchange and digital signatures, but the urgent question for adopters is how to operationalise PQC in complex environments with confidence. In banking, Transport Layer Security (TLS), for example, protects data‑in‑transit across public‑facing channels and internal services, and is terminated at many heterogeneous endpoints (web servers, API gateways, load balancers, reverse proxies), each a potential quantum‑vulnerable component and migration target.

We argue that the bottleneck is operational rather than algorithmic: hybrid key exchanges such as X25519‑ML‑KEM‑768 are already available in mainstream libraries, but security teams lack precise visibility into TLS configurations and repeatable methods for enabling PQC‑compatible settings across a heterogeneous estate. This paper presents a configuration‑parsing methodology that automatically extracts and normalises TLS cryptographic posture across dominant enterprise web‑server stacks, producing a unified, provenance‑traced cryptographic inventory as a foundation for migration and compliance. We demonstrate the approach on 8,443 real‑world Nginx configurations from public repositories and in a proof‑of‑concept deployment at a financial institution, where ML‑KEM‑512 and X25519‑ML‑KEM‑768 are onboarded at TLS termination points (web server and API gateway) securing an internal application, with zero application‑layer changes and manageable performance overhead.

\end{abstract}


\section{Introduction}
\label{sec:intro}

Organisations around the world are undertaking a massive transformation of their cryptographic infrastructure in order to become quantum-safe, before large-scale quantum computers become a reality. This threat is well understood, where a cryptographically relevant quantum computer (CRQC) running Shor's algorithm~\cite{shor1994} would break the mathematical hardness assumptions underlying RSA and elliptic-curve cryptography (ECC), which serve as the foundations of virtually every public-key cryptographic scheme in use today for security in the internet. Beyond this future threat, the \textit{harvest-now, decrypt-later} (HNDL) attack model poses an immediate risk: well-resourced adversaries can already intercept and archive encrypted traffic today, with the intention of decrypting it retroactively, once a quantum computer large enough to break RSA/ECC is available~\cite{tno2024handbook}. This is especially relevant for organisations handling sensitive data with long-term confidentiality requirements (e.g.) financial records, legal instruments, healthcare data. We are witnessing continuous improvements in the required number of qubits and quantum resources to break classical public-key cryptography, with the most recent work of Babbush et al., demonstrating improved quantum algorithms and implementation techniques significantly reduce the estimated resources needed to break elliptic‑curve cryptocurrencies, tightening the urgency of timely PQC migration~\cite{babbush2026securing}.

Standardisation bodies and government agencies around the world have been preparing to counter this threat for more than a decade now. The standardisation of PQC schemes by NIST in 2024 really moved the needle, towards widely adopting PQC in real-world systems. In August 2024, NIST finalized ML-KEM (FIPS~203)~\cite{nist_fips203} for key encapsulation, ML-DSA (FIPS~204)~\cite{nist_fips204} for digital signatures, and SLH-DSA (FIPS~205)~\cite{nist_fips205} as a conservative hash-based signature alternative. There are more schemes that are planned to be standardized not just by NIST, but also by other standardisation bodies around the world in different countries, in a bid to have multiple options for quantum security. 
These standards address the algorithmic dimension of the problem. The urgent question that now faces organisations that want to migrate is however a different one, and is more focussed on operations: \textit{how do I operationalise PQC in my environment with confidence?} The question mainly revolves around migration playbooks that are repeatable, and those that rely on sound methodologies that operational teams managing infrastructure on a day-to-day basis can use, to turn on the PQC feature in their systems with full confidence.

\subsection*{The Operational Challenge in Financial Infrastructure}

Transport Layer Security (TLS) is the most widely used security protocol in banking environments, securing data-in-transit across public-facing digital channels, internal service-to-service
communication, API calls, and several other use cases. Modern financial infrastructures rely on TLS termination at dozens to hundreds of heterogeneous endpoints, that are present across various modern service gateways such as web servers, API gateways, load balancers, reverse proxies, database servers and many more. Each of these components and their instances are typically independently configured, each is a potential quantum-vulnerable component, and each is a migration target.

The challenge of onboarding PQC into this environment is not about the algorithm itself. For instance, there is support for hybrid key exchange schemes such as X25519-MLKEM768 in current releases of popular cryptographic libraries such as OpenSSL and BouncyCastle~\cite{oqs_provider, bouncycastle}. The main challenge however lies in the operational aspects: gaining precise visibility into what each termination point is actually configured to use, and having a sound and repeatable methodology for enabling PQC-compatible configurations across a heterogeneous infrastructure estate. This is even more important, considering the possibility of need for more frequent cryptographic upgrades in the future, with constant advancements in computing capabilities.

Specifically focussing on TLS, an organisation cannot execute a carry out a cryptographic upgrade across its TLS endpoints with confidence unless it can first answer, with certainty, what cryptographic algorithms and parameters each of its TLS endpoints is currently running, not what they negotiate under observed conditions, but all the allowed configurations that are permitted to be used. More so, it is important to understand how different systems can be configured to use certain types of cryptographic algorithms, and have a way to automatically infer how they are configured to operate. While there are approaches such as network-level scanners, they observe negotiated handshakes and cannot reveal permitted, but fallback cipher suites, uninvoked legacy protocols, or HSM-backed key references that are declared in configuration but never exercised under normal traffic. This paper mainly addresses this visibility gap as the central operational problem.

\subsection*{Our Contributions}

We present a novel configuration-parsing methodology for automated extraction and normalisation of TLS cryptographic posture across the dominant web server technologies used in enterprise environments. The methodology produces a unified, provenance-traced cryptographic inventory that gives security teams an exact picture of every protocol version, cipher suite, certificate reference, and HSM dependency across their infrastructure, as the foundation for migration planning and compliance verification.

We make three concrete contributions.

\begin{itemize}

  \item[\textbf{C1.}] \textbf{Configuration-Parsing Methodology for TLS Discovery.}
  We design and implement a deterministic, config-level TLS extraction and normalisation framework that is applicable to dominant web server technologies in enterprise environments. The framework parses vendor-specific configuration files, resolves inheritance semantics, and normalises
  the output into a unified, technology-agnostic representation with full provenance tracing. It supports automated policy comparison against NIST SP~800-52, PCI-DSS, CIS Benchmarks, and custom
  quantum-readiness policies~\cite{nist_sp80052}, giving security teams a repeatable, auditable method for assessing and improving TLS posture across heterogeneous infrastructure.

  \item[\textbf{C2.}] \textbf{Proof-of-Concept PQC Deployment in a
  Financial Institution.} We also demonstrate the practical effectiveness of the proposed approach, by implementing it within a proof-of-concept environment at a financial institution, where we onboarded quantum-safe key exchange schemes such as ML-KEM-512 and X25519-MLKEM768 at production TLS termination points such as public-facing web servers and API gateways. These TLS termination points were used to deliver data to an internal banking application. We were able to enable PQC in the web server and API gateways, and notably did not require any changes to the application itself. We further also report end-to-end performance benchmarks under 100 concurrent
  clients, demonstrating manageable overhead and zero errors across all test runs.

  \item[\textbf{C3.}] \textbf{Large-Scale Measurement in the Wild.}
  We also expand our approach, and apply it to open-source projects in the wild, and performed an experiment on a corpus of 8,443 real-world Nginx configurations drawn from public repositories. This study provides an empirical baseline of TLS hygiene and quantum vulnerability surface in the wild, covering protocol adoption, cipher suite hygiene, key exchange vulnerability, certificate management, and configuration drift across actively maintained and archived repositories.

\end{itemize}

Together, the aforementioned contributions (C1-C3) provide a practical, evidence-driven path for
operationalising PQC in enterprise TLS infrastructure, from gaining visibility over the current cryptographic estate, to demonstrating a concrete migration in a real financial environment, to establishing how the broader open-source ecosystem stands today.

\subsection*{Paper Organisation}

The remainder of this paper is organised as follows. Section~\ref{sec:background} provides background on the quantum threat, TLS as a migration surface, and relevant related work.
Section~\ref{sec:system} describes the configuration-parsing methodology for TLS discovery and normalisation. Section~\ref{sec:measurement} presents the large-scale measurement study of open-source Nginx configurations. Section~\ref{sec:casestudy} details the PQC deployment at the
financial institution. Section~\ref{sec:discussion} synthesises findings into a migration framework and discusses limitations. Section~\ref{sec:conclusion} provides conclusion to the paper and briefly touches upon future work. 

\section{Background and Related Work}
\label{sec:background}

\subsection{The Quantum Threat to Public-Key Cryptography}

The security of RSA and elliptic-curve Diffie-Hellman (ECDH) rests on
the presumed hardness of integer factorization and the discrete logarithm
problem, respectively. In 1994, Shor demonstrated that a quantum computer
can solve both problems in polynomial time~\cite{shor1994}, rendering
any system whose long-term security depends on these primitives vulnerable
once sufficiently large quantum hardware becomes available. Grover's
algorithm~\cite{grover1996} provides a quadratic speedup for unstructured
search, effectively halving the security level of symmetric primitives:
AES-128 is reduced to approximately 64-bit security against a quantum
adversary, while AES-256 retains approximately 128-bit security. The
practical implication is public-key or asymmetric cryptography requires
replacement, while private-key or symmetric cryptography requires only key-size increases.

Current consensus places the timeline for a cryptographically relevant
quantum computer (CRQC) at somewhere between ten and thirty years, though this estimate carries substantial uncertainty~\cite{tno2024handbook,postquantum2025enterprise}. The HNDL
threat model, however, decouples migration urgency from CRQC arrival:
data encrypted today under quantum-vulnerable algorithms can be archived
and decrypted retroactively. For financial records, legal instruments,
and government communications with multi-decade confidentiality
requirements, this makes the migration imperative immediate rather than
deferred~\cite{tno2024handbook}.

\subsection{NIST Post-Quantum Cryptography Standardization}

Following an eight-year public evaluation process, NIST selected four algorithms for standardisation in 2022, and finalized three
post-quantum cryptographic standards in August 2024. ML-KEM
(FIPS~203)~\cite{nist_fips203}, derived from CRYSTALS-Kyber, specifies
a module-lattice-based key encapsulation mechanism providing IND-CCA2
security under the Module Learning With Errors (MLWE) hardness
assumption. ML-DSA (FIPS~204)~\cite{nist_fips204}, derived from
CRYSTALS-Dilithium, provides a lattice-based digital signature scheme.
SLH-DSA (FIPS~205)~\cite{nist_fips205}, derived from SPHINCS+, provides
a stateless hash-based signature scheme whose security rests solely on
hash function properties. HQC was also recently selected for standardisation in 2025 by NIST, and there will be more algorithms that will be standardised for digital signatures with specific properties and use-cases.

For TLS key exchange, the primary quantum-vulnerable component in
most enterprise deployments, ML-KEM is the relevant primitive. It
replaces ECDH in the TLS~1.3 key exchange~\cite{rfc8446}, and is
deployed in practice as a hybrid scheme combining classical ECDH
(e.g., X25519) with ML-KEM (e.g., X25519-MLKEM768), providing
security against both classical and quantum adversaries simultaneously.
Hybrid deployment is recommended during the transition period to guard
against unforeseen weaknesses in the new algorithms~\cite{nist_cswp39}.

\subsection{TLS as the Primary Migration Surface}

TLS~1.3~\cite{rfc8446} is the dominant secure transport protocol for
enterprise communications. Its key exchange is based on ephemeral
Diffie-Hellman (ECDHE or DHE), which is quantum-vulnerable via Shor's
algorithm. Digital certificates used for authentication rely on RSA or
ECDSA signatures, which are similarly vulnerable, though certificate
lifetimes are shorter and the migration path is distinct. TLS~1.0
and~1.1 were formally deprecated by RFC~8996~\cite{rfc8996} and are
banned under PCI-DSS~v4.0, yet as we show in
Section~\ref{sec:measurement}, they persist in a significant fraction
of real-world configurations.

The challenge of TLS migration in enterprise environments is
fundamentally an \textit{infrastructure perimeter} problem. In modern
financial institutions, TLS termination is not centralized but
distributed: load balancers, CDN edge nodes, API gateways, reverse
proxies, and application-layer web servers each independently terminate
TLS sessions, negotiate cryptographic parameters, and manage
certificates and private keys. This distributed termination model is
operationally necessary, it enables horizontal scaling, traffic
routing, and defence-in-depth, but it means that a financial
institution may operate dozens to hundreds of independent TLS
termination points, each a potential quantum-vulnerable component.
Critically, this perimeter secures the institution's most sensitive
systems: core banking platforms, interbank payment channels, fraud
detection pipelines, and customer-facing digital banking services.
A failure to migrate any one of these termination points leaves a gap
through which HNDL-collected traffic can eventually be decrypted.

The heterogeneity of TLS implementations across this
perimeter, Nginx on public-facing web servers, Spring Boot on
internal microservices, Java-based API gateways, hardware load
balancers with proprietary TLS stacks, creates significant operational
complexity for migration. Each technology exposes TLS configuration
through a different interface, uses different directive names, and
applies different inheritance semantics. Network-level scanners cannot
address this complexity because they observe \textit{negotiated}
behavior, not \textit{declared} configuration: a server may negotiate
TLS~1.3 with a strong cipher suite in every live handshake while its
configuration file still lists TLS~1.0 and weak cipher suites as
permitted fallbacks, a ticking clock invisible to any scanner. This
fragmentation, and this visibility gap, are the core motivations for
the automated discovery framework described in Section~\ref{sec:system}.

\subsection{Related Work}

\paragraph{TLS Measurement Studies.}
Large-scale empirical analyses of TLS deployment have a rich history.
Holz et al.\ conducted one of the earliest systematic studies of
TLS in the wild, examining certificate and protocol adoption across
the public web. More recently, Hebrok et al.~\cite{hebrok2023tls}
analyzed TLS session ticket cryptography across millions of servers,
identifying widespread cryptographic weaknesses in session resumption.
van den Berg et al.~\cite{vanderburg2020https} studied HTTPS deployment
practices at scale, finding persistent adoption of deprecated
configurations. Our measurement study differs from these in two
important respects: we operate at the \textit{configuration file} level
rather than the network handshake level, exposing permitted-but-uninvoked
cipher suites and declared fallback policies; and we explicitly quantify
the quantum vulnerability surface of real-world deployments, an angle
absent from prior work.

\paragraph{PQC Integration in TLS.}
Paquin et al.\ evaluated the performance of post-quantum and hybrid key
exchange in TLS, providing early benchmarks for Kyber and other NIST
candidates. Stebila and Sullivan proposed a framework for hybrid key
exchange in TLS~1.3, which forms the basis for the IETF draft on hybrid
key exchange design. Most recently, the Looma system~\cite{looma2026}
proposed a low-latency PQTLS authentication architecture for cloud
environments. Our work complements these by focusing on the
\textit{deployment} and \textit{operational} dimensions of PQC migration, rather than protocol design or
algorithmic performance in isolation.


\paragraph{PQC Migration Guidance.}
The TNO PQC Migration Handbook~\cite{tno2024handbook} provides the most
comprehensive practitioner guidance on migration planning, covering
cryptographic inventory, risk prioritization, and migration sequencing.
Our work is empirically grounded where these documents are prescriptive:
we measure the gap between current practice and quantum-safe posture, and
we report a concrete deployment case in a financial environment.

\section{Automated TLS Discovery and Normalization}
\label{sec:system}

\subsection{Motivation and Design Goals}

One of the primary questions facing organisations that want to migrate to post-quantum cryptography is: \textit{what cryptography is each component of my infrastructure currently configured to use?} This question is harder than it appears. A large financial institution operates a diverse technology estate assembled from hardware and software appliances across multiple vendors, each with its own
management interface, configuration model, and operational semantics. 
For instance, a typical organisation runs \texttt{nginx} instances as public-facing reverse proxies, \texttt{Apache} web servers hosting internal applications, Spring Boot microservices behind API gateways, hardware and software load balancers from different vendors, and database servers with their own TLS stacks. Each of these components controls its cryptographic behaviour differently: different configuration syntax, different directive naming, different inheritance models, and different vendor-specific defaults. There exists no common tooling today, that can query all of these components and return a consistant, comparable answer about their cryptographic posture, as well have the ability to configure them to behave according to a specific set of cryptographic requirements.

In this work, we focus specifically on software-based web servers and API gateways, which expose their cryptographic configuration through human-readable configuration files. This scope is practically significant because these components are the primary TLS termination points in most enterprise architectures, sitting directly in front of core applications and services that depend on them for secure communication. Hence, they also serve as high priority migration targets. They are also technical tractable, since configuration files provide a deterministic, parseable source of truth about what cryptographic algorithms and parameters a component is permitted to use, as well as have the ability to control and configure it as well.


There are a few other approaches to determine cryptographic posture, with a prominent method being use of network-level scanners. Network-level scanners such as Qualys SSL Labs and \texttt{testssl.sh} probe the TLS handshake: they observe what a server \textit{negotiates} under normal conditions. This is insufficient for two reasons. First, a server may negotiate TLS~1.3 with a modern cipher suite in every observed handshake while its configuration file permits TLS~1.0 as a fallback, an attack surface that is real but invisible to the scanner. Second, scanners cannot inspect important aspects such as HSM-backed key references, certificate chain completeness, or session management policy, all of which are critical inputs to migration planning.

We therefore design a framework that operates directly on the configuration files used to control and configure TLS termination points, with four explicit goals:

\begin{enumerate}
  \item \textbf{Determinism.} Every extracted value must be traceable
    to a specific directive in a specific file. Probabilistic or
    inferred values are unacceptable in a security-critical inventory.
  \item \textbf{Normalization.} The same TLS concept expressed differently
    across different stacks must product identical output, thereby enabling cross-technology comparison. In this work, we specifically look at \texttt{nginx}, \texttt{Apache} and \texttt{Spring Boot} based stacks.
  \item \textbf{Provenance.} Every field in the output must carry a
    \texttt{derived\_from} trace recording the exact details from the configuration files such as 
    the original directive name, raw value, and source file location, so that every compliance
    violation can be linked back to the exact configuration line
    that requires remediation.
  \item \textbf{Extensibility.} Adding support for a new technology or
    a new TLS directive must require only a declarative mapping change,
    not modification of core parsing logic.
\end{enumerate}

\subsection{System Architecture}

\begin{figure*}[!t]
    \centering
    \includegraphics[width=0.8\linewidth]{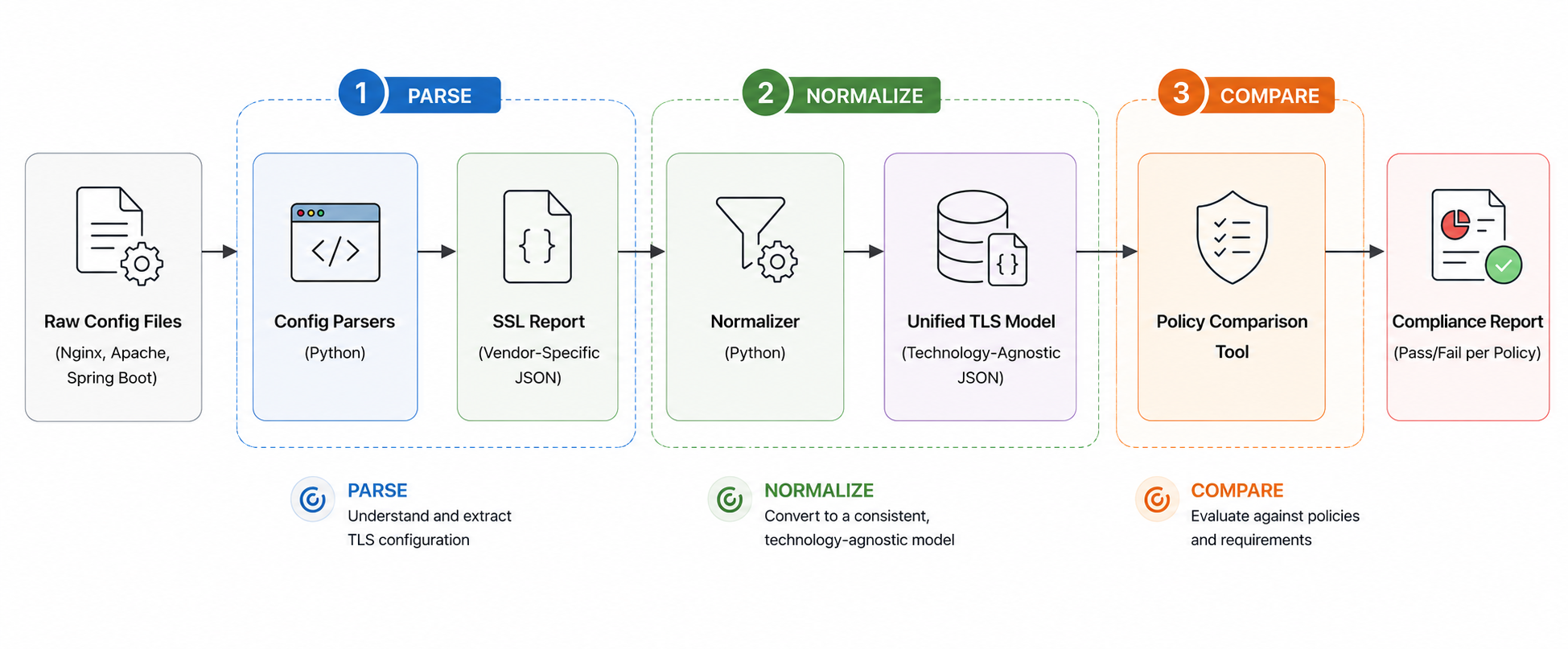}
    \caption{The three-stage pipeline: \textsc{Parse}, \textsc{Normalize}, and \textsc{Compare}.}
    \label{fig:three_stage_pipeline_python}
\end{figure*}

The framework follows a three-stage pipeline: \textsc{Parse},
\textsc{Normalize}, and \textsc{Compare}. Figure~\ref{fig:three_stage_pipeline_python}
illustrates the overall architecture.

\subsubsection{Stage 1: Technology-Specific Parsing}

Each supported technology has a dedicated parser that understands its
configuration grammar and inheritance model.

\paragraph{\texttt{nginx}.} \texttt{nginx} configuration follows a strict block-directive
grammar with a well-defined inheritance chain: global scope flows into
\texttt{http} blocks, which flow into \texttt{server} blocks. Each
\texttt{server} block that includes \texttt{listen 443 ssl} constitutes
an independent TLS context. The parser uses the \texttt{crossplane}
library to construct a syntax tree, then walks it to resolve inheritance
and extract TLS-relevant directives per server block.

\paragraph{\texttt{apache} HTTP Server.} The configuration of \texttt{apache} is governed by
a set of directives with defined scoping rules across \texttt{<VirtualHost>} and \texttt{<Directory>} containers. The parser uses the \texttt{apacheconfig} library and applies Apache's precedence rules to determine the effective value of each TLS directive per virtual
host.

\paragraph{\texttt{Spring Boot.}} \texttt{Spring Boot} exposes TLS configuration through
YAML or properties files, potentially split across multiple profile
documents that are merged at runtime. The parser uses the \texttt{pyYaml}
library and implements \texttt{Spring Boot}'s profile-merging semantics to produce
a single effective configuration per application deployment.

All parsers share a common base class that enforces the extraction
workflow: read the raw file with size bounds; apply the
technology-specific parsing library; resolve inheritance; and emit a
structured intermediate JSON retaining native directive terminology.
Only TLS-relevant directives defined in a property allowlist are
extracted, covering the ten directive categories whose misconfiguration
most commonly leads to security exposure: protocol versions, cipher
suites, certificate and key paths, client authentication settings,
trust store references, session management, OCSP stapling, HSTS
headers, cipher order preference, and ECDH curve selection.

\subsubsection{Stage 2: Normalization into the Unified TLS Model}

The intermediate vendor-specific output is passed to a universal
normalizer, which transforms it into the \textit{Unified TLS Model}
(UTM), a technology-agnostic schema representing TLS configuration
regardless of origin. The UTM organizes configuration into eight
logical sections:

\begin{itemize}
  \item \textbf{Endpoint:} port, hostname, server role
  \item \textbf{Protocols:} enabled TLS version set
  \item \textbf{Ciphers:} permitted cipher suite list, server-order
    preference flag, ECDH curve selection
  \item \textbf{Certificates:} leaf certificate path, private key
    path, OCSP stapling configuration
  \item \textbf{Trust:} CA certificate paths, CRL paths
  \item \textbf{Verification:} client authentication mode, verify
    depth, strict SNI enforcement
  \item \textbf{Session:} session cache type and timeout, ticket
    support, compression
  \item \textbf{Security Headers:} HSTS configuration
\end{itemize}

Normalization is driven by declarative mapping files, one per
technology, that specify, for each vendor directive, the transformation
to apply and the UTM field to populate. Crucially, every UTM field
carries a \texttt{derived\_from} record containing the original
directive name and raw value, preserving the full provenance chain
from normalized output back to source file.

\subsubsection{Stage 3: Policy Comparison}

The UTM is evaluated against a configurable set of TLS security
policies. Each policy defines per-field requirements; the comparison
engine produces a structured pass/fail report per server context per
policy. Supported policy profiles include NIST SP~800-52
Rev.~2~\cite{nist_sp80052}, PCI-DSS~v4.0, CIS Benchmarks for Nginx
and Apache, Mozilla Modern and Intermediate profiles, and custom
organizational policies. A quantum-readiness policy decomposes each
cipher suite into its constituent algorithms and flags any component
vulnerable to Shor's or Grover's algorithms, enabling automated
classification of the entire cryptographic inventory by quantum risk
level.

\subsection{HSM Awareness}

In high-security environments such as banking and critical
infrastructure, private keys are stored in Hardware Security Modules
(HSMs), dedicated, tamper-resistant devices that generate and manage
keys without exposing them to the host operating system. The framework
detects HSM-backed configurations by identifying PKCS\#11 URI
references and engine directives in place of filesystem key paths.
This distinction is critical for migration planning: HSM firmware may
require updates to support PQC algorithms, and HSM-backed systems
have different key rotation procedures and migration timelines than
filesystem-backed ones. Identifying HSM-backed systems before migration
begins prevents the common failure mode of discovering HSM
incompatibilities mid-migration.

\subsection{Migration Workflow Integration}

The framework supports a five-phase migration lifecycle: (1)
\textbf{Discover}, build the cryptographic inventory from parsed
configurations; (2) \textbf{Assess}, apply quantum-readiness policies
to classify every cipher suite and key reference by risk; (3)
\textbf{Prioritize}, rank migration targets by exposure, data
sensitivity, and HSM constraints; (4) \textbf{Migrate}, deploy
hybrid PQC configurations and re-parse to confirm changes; (5)
\textbf{Verify}, run policy comparison against a post-quantum policy
to confirm removal of vulnerable primitives and detect regressions.
This loop is continuous: the same pipeline that builds the initial
inventory becomes the ongoing monitoring layer for the post-migration
environment.

\section{The State of TLS in the Wild: A Large-Scale Measurement Study}
\label{sec:measurement}

\subsection{Dataset Construction}

To establish an empirical baseline of real-world TLS configuration
practices, we collected and analyzed Nginx configuration files from
public GitHub repositories. We focus on Nginx for three reasons: its
explicit configuration model exposes every cryptographic decision as
a visible, greppable directive; its hierarchical inheritance model
allows precise per-virtual-host TLS posture determination; and its
ubiquity in public repositories yields a large, naturally occurring
dataset of real-world cryptographic choices.

\paragraph{Collection Pipeline.}
The collection pipeline operates in four stages. First, we enumerate
TLS-relevant Nginx directives from our property allowlist and pair
each with known Nginx configuration filenames (\texttt{nginx.conf},
\texttt{default.conf}, \texttt{ssl.conf}), producing a matrix of
search queries. Second, each query is executed against the GitHub
Code Search API; results are deduplicated by owner, repository, and
file path. Third, raw file content is fetched via the GitHub Contents
API; a SHA-256 hash is computed per file for cross-run deduplication.
Fourth, each file is passed through a confidence classifier that
evaluates structural markers and assigns a score in $[0,1]$; files
below a threshold are excluded.

\paragraph{Active vs.\ Archived Classification.}
For every collected file, we query the GitHub API for two signals:
the repository's \texttt{archived} flag and the last-push date
(repositories with no push activity in over two years are classified
as dormant). This distinction enables comparison of TLS practices
between actively maintained configurations and those frozen at a
prior point in time.

\paragraph{Dataset Summary.}
The pipeline yielded 9,183 manifest entries, resolving to 8,443
unique configuration files after deduplication: 5,524 from active
repositories and 2,919 from archived repositories. Parsing produced
12,875 server-block contexts; of these, 5,982 had TLS enabled and
form the core dataset for all subsequent analysis.

\subsection{Research Questions and Findings}

\subsubsection{RQ1: Protocol Adoption}

\begin{table}[t]
  \centering
  \caption{TLS Protocol Version Adoption (5,982 TLS-Enabled Contexts)}
  \label{tab:protocols}
  \begin{tabular}{lc}
    \toprule
    \textbf{Protocol}          & \textbf{Adoption Rate} \\
    \midrule
    TLS 1.2                    & 95.8\%                 \\
    TLS 1.3                    & 73.5\%                 \\
    TLS 1.1                    & 21.6\%                 \\
    TLS 1.0                    & 19.4\%                 \\    
    \bottomrule
  \end{tabular}
\end{table}

TLS~1.2 is near-universal at 95.8\%, and TLS~1.3 adoption is strong
at 73.5\%. However, 21.8\% of configurations still permit TLS~1.0
or~1.1, both formally deprecated by RFC~8996~\cite{rfc8996} and
prohibited by PCI-DSS~v4.0.

\begin{figure}[h]
    \centering
    \includegraphics[width=\linewidth]{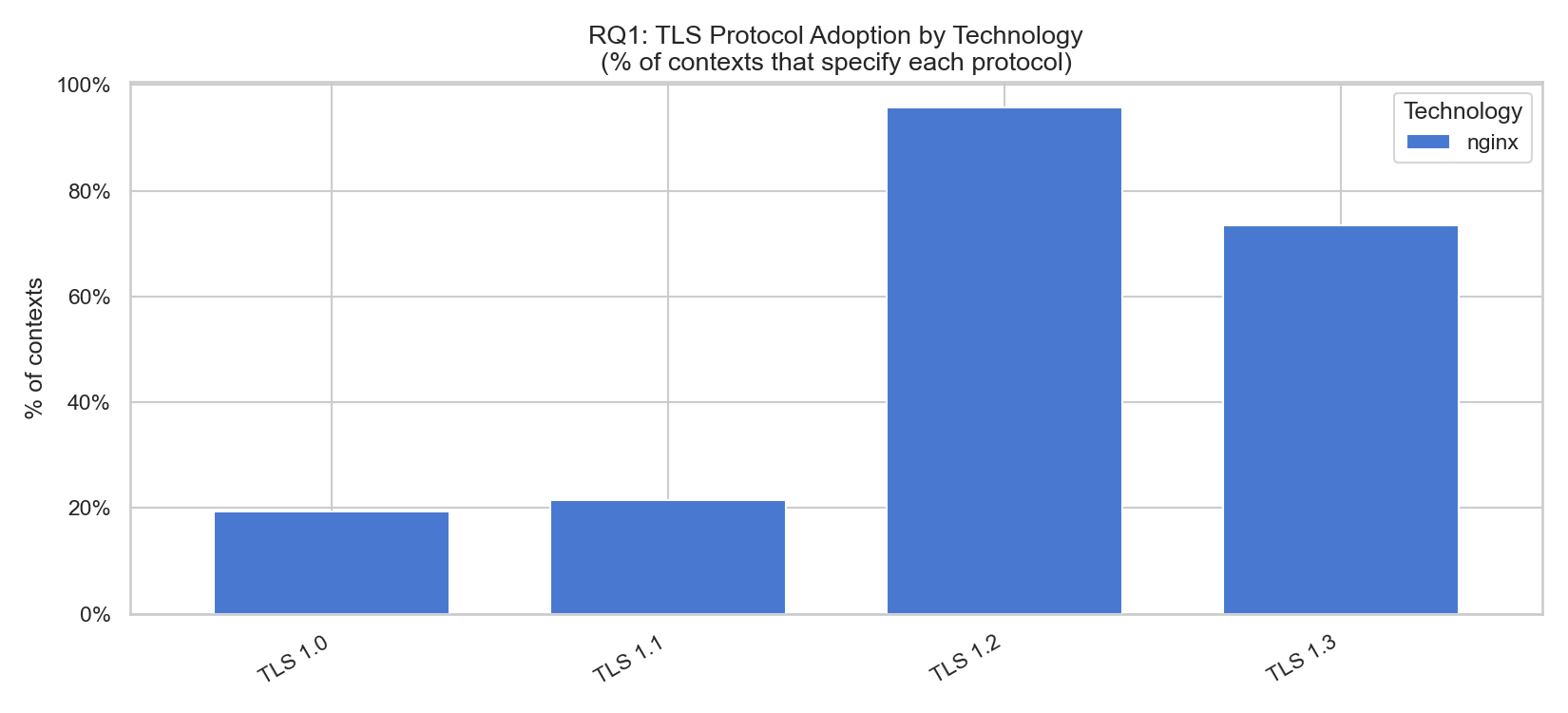}
    \caption{Protocol Adoption}
    \label{fig:rq1_protocol_adoption.png}
\end{figure}

Figure~\ref{fig:rq1_protocol_adoption.png}.

\subsubsection{RQ2: Cipher Suite Hygiene}

Of the 3,603 TLS-enabled contexts with explicit cipher
configurations, 37.6\% include at least one weak cipher token
(RC4, DES, 3DES, EXPORT, NULL, or MD5, excluding properly negated
entries). The most frequently co-occurring cipher pair is
\texttt{ECDHE-RSA-AES128-GCM-SHA256} and
\texttt{ECDHE-RSA-AES256-GCM-SHA384}, appearing together in 1,737
contexts, indicating strong AEAD adoption in the well-maintained
segment. The weak cipher tail reflects copy-paste from outdated
sources without security audit.

\begin{figure}[h]
    \centering
    \includegraphics[width=\linewidth]{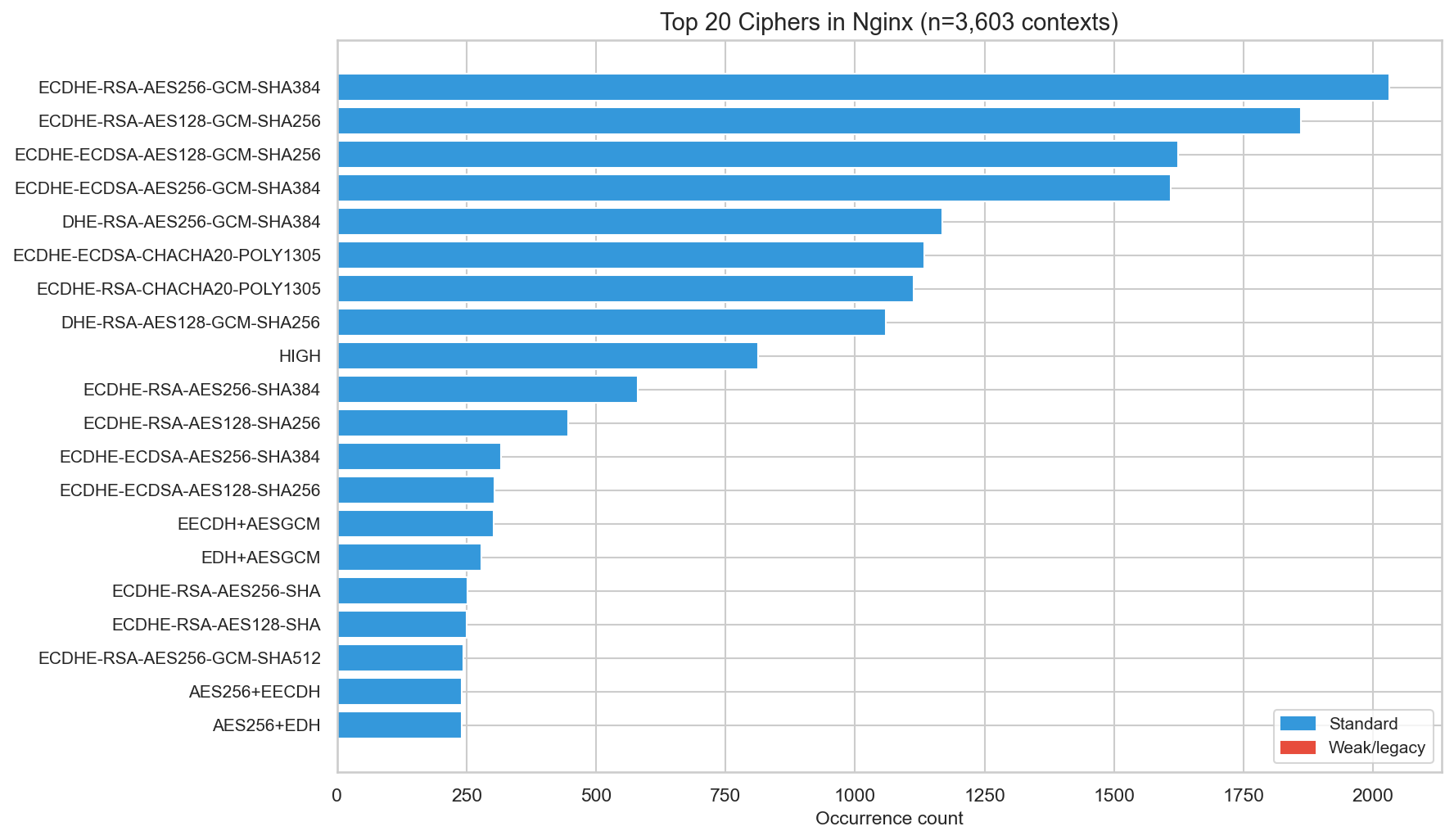}
    \caption{Cipher Suite Hygiene}
    \label{fig:rq2_ciphers_nginx.png}
\end{figure}

Figure~\ref{fig:rq2_ciphers_nginx.png}.

\begin{figure}[h]
    \centering
    \includegraphics[width=\linewidth]{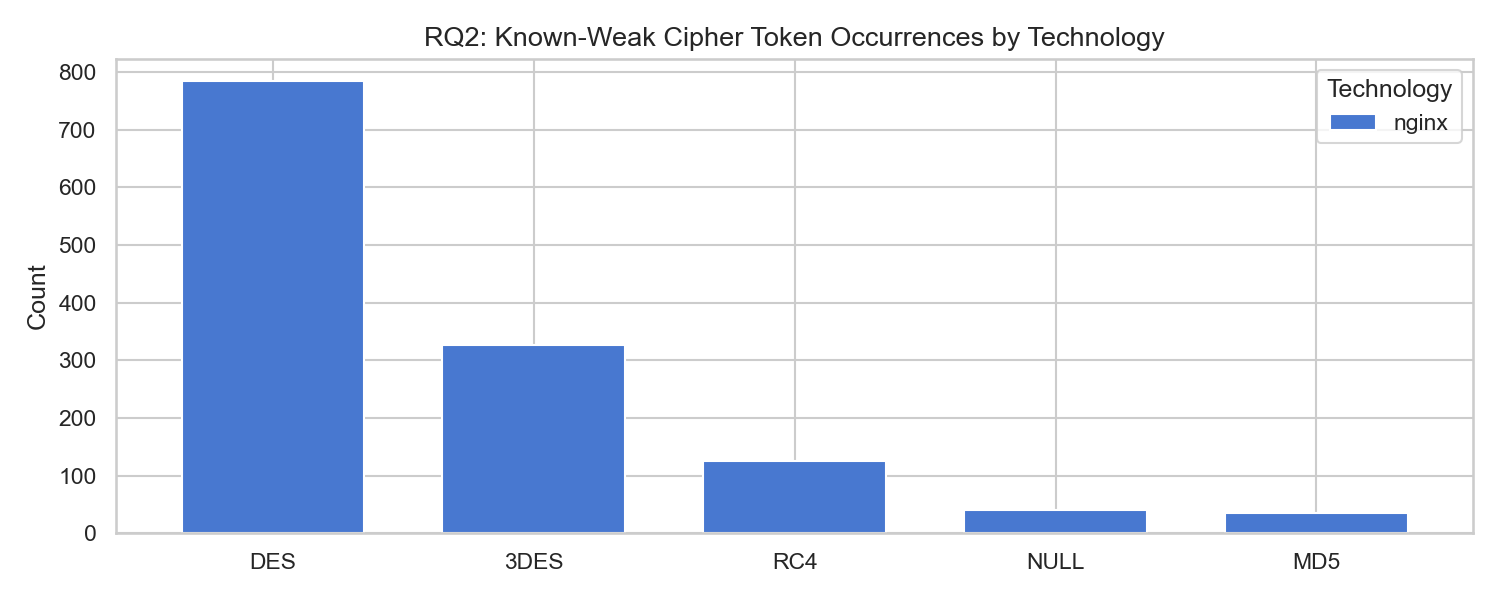}
    \caption{Cipher Suite Hygiene}
    \label{fig:rq2_known_weak_cipher_token_occurences_by_tech.png}
\end{figure}

Figure~\ref{fig:rq2_known_weak_cipher_token_occurences_by_tech.png}.

\subsubsection{RQ3: Quantum Vulnerability Surface}

\begin{table}[t]
  \centering
  \caption{Key Exchange Categories by Quantum Vulnerability}
  \label{tab:quantum}
  \begin{tabular}{lc}
    \toprule
    \textbf{Key Exchange Category}                  & \textbf{Share} \\
    \midrule
    ECDHE/DHE (forward secrecy, quantum-vulnerable) & 71.1\%         \\
    RSA KE (no forward secrecy, quantum-vulnerable) & 28.9\%         \\
    PQC hybrid (quantum-resistant)                  & 0.0\%          \\
    \bottomrule
  \end{tabular}
\end{table}

\begin{figure}[h]
    \centering
    \includegraphics[width=\linewidth]{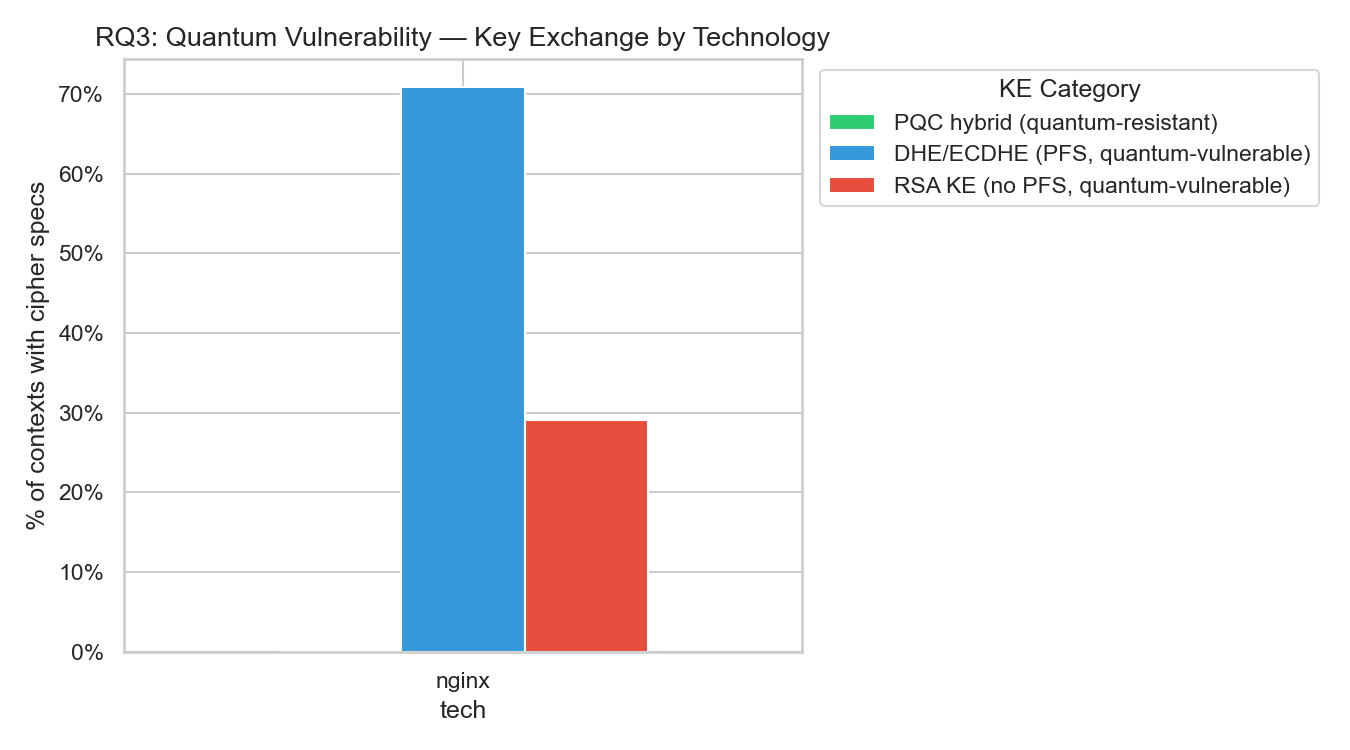}
    \caption{Quantum Vulnerability Surface}
    \label{fig:rq3_quantum_vulnerability.png}
\end{figure}

Post-quantum hybrid key exchange adoption is zero in this corpus.
The RSA key exchange category (28.9\%) is particularly concerning:
it provides no forward secrecy, meaning compromise of the server's
private key, whether by a classical adversary today or a quantum
adversary in the future, retroactively decrypts all previously
recorded sessions. Among the 625 contexts explicitly configuring
\texttt{ssl\_ecdh\_curve}, the post-quantum hybrid curve
\texttt{X25519MLKEM768} appears in exactly four configurations, 0.8\%
of curve-specifying contexts, representing the earliest adopters
of post-quantum key exchange in this dataset. Figure~\ref{fig:rq3_quantum_vulnerability.png}.

\subsubsection{RQ4: Certificate and Key Management}

\begin{figure}[h]
    \centering
    \includegraphics[width=\linewidth]{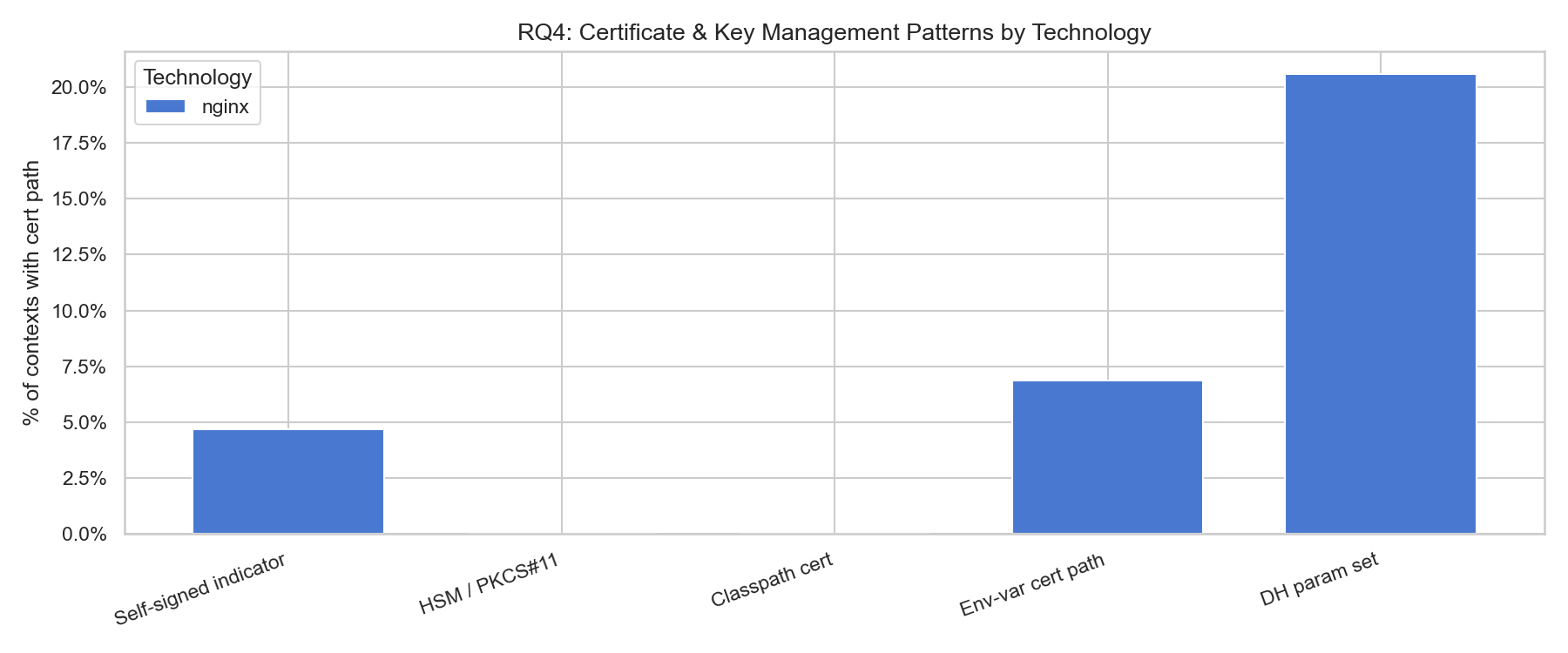}
    \caption{Certificate and Key Management}
    \label{fig:rq4_cert_management.png}
\end{figure}

Certificate and key paths are present in 95.0\% and 96.1\% of
contexts respectively. DH parameters are configured in only 20.6\%
of contexts. HSM/PKCS\#11 references are absent from the entire
dataset, consistent with the expectation that hardware-backed key
protection is not represented in public repositories.
Environment-variable certificate paths (6.9\%) reflect
container-native deployments. Figure~\ref{fig:rq4_cert_management.png}.

\subsubsection{RQ5: HSTS Adoption}

\begin{figure*}[h]
    \centering
    \includegraphics[width=0.75\linewidth]{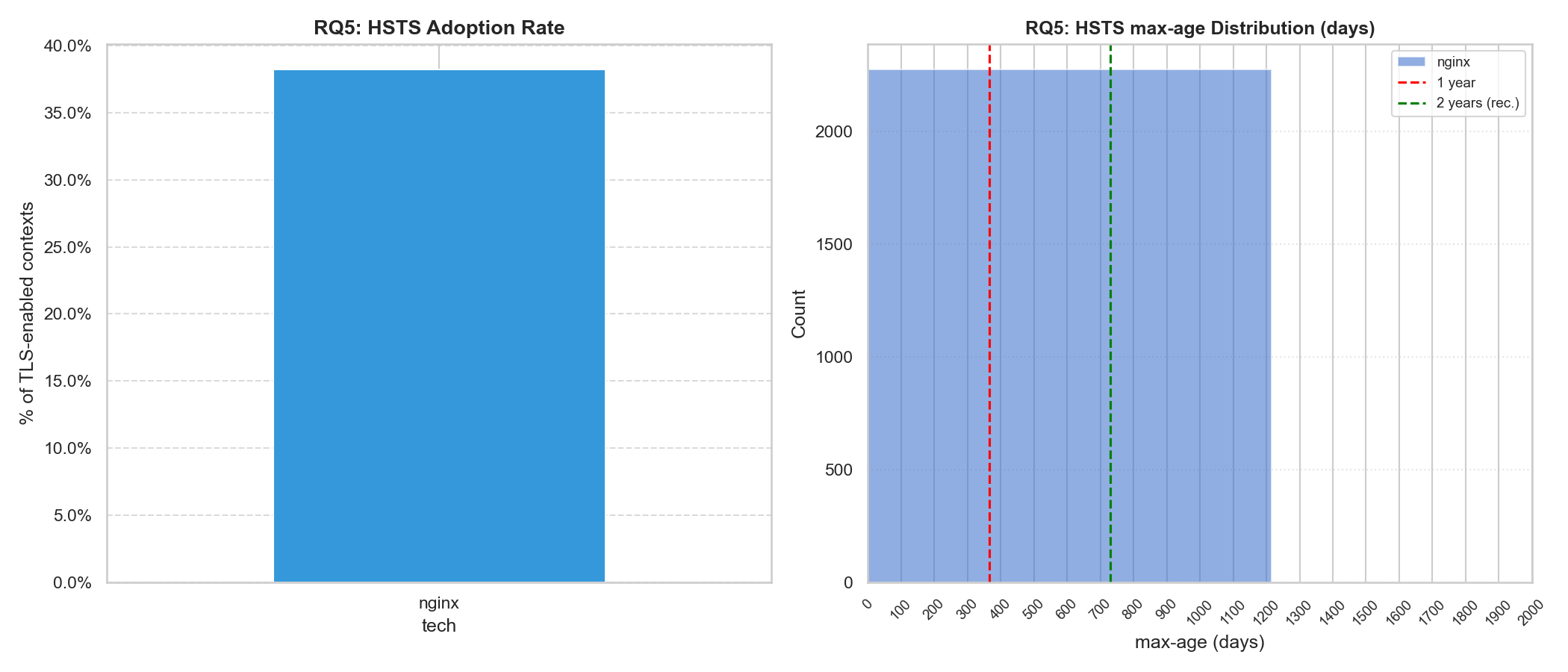}
    \caption{HSTS Adoption}
    \label{fig:rq5_hsts.png}
\end{figure*}

Only 38.2\% of TLS-enabled contexts configure HSTS. Nearly
two-thirds of servers that established TLS still leave the door open
to SSL stripping attacks a network-level interception technique
that HSTS prevents at zero performance cost. HSTS is arguably the
highest-impact, lowest-effort transport security control, and its
absence from the majority of configurations reflects a systemic gap
between TLS deployment and transport security posture. Figure~\ref{fig:rq5_hsts.png}.

\begin{figure*}[h]
    \centering
    \includegraphics[width=0.75\linewidth]{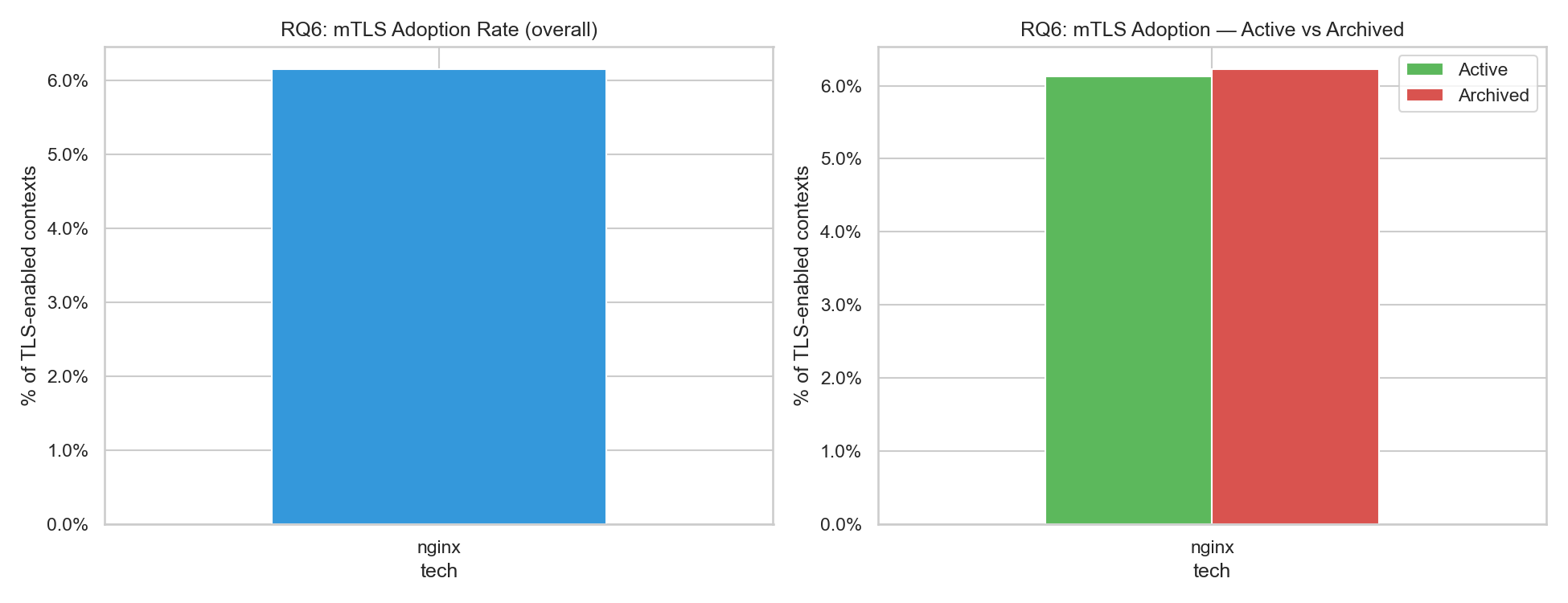}
    \caption{Mutual TLS (mTLS)}
    \label{fig:rq6_mtls.png}
\end{figure*}

\subsubsection{RQ6: Mutual TLS}

mTLS adoption is 6.2\% overall — low, but expected given that most public-facing web servers authenticate clients via application-layer mechanisms (cookies, tokens, OAuth) rather than client certificates. The rate is virtually identical between active (6.1\%) and archived (6.2\%) repos, suggesting mTLS adoption is driven by architectural choice (service mesh, internal APIs) rather than security trend evolution. Figure~\ref{fig:rq6_mtls.png}.

\subsubsection{RQ7: Mixed-Strength Configurations}

\begin{figure}[h]
    \centering
    \includegraphics[width=\linewidth]{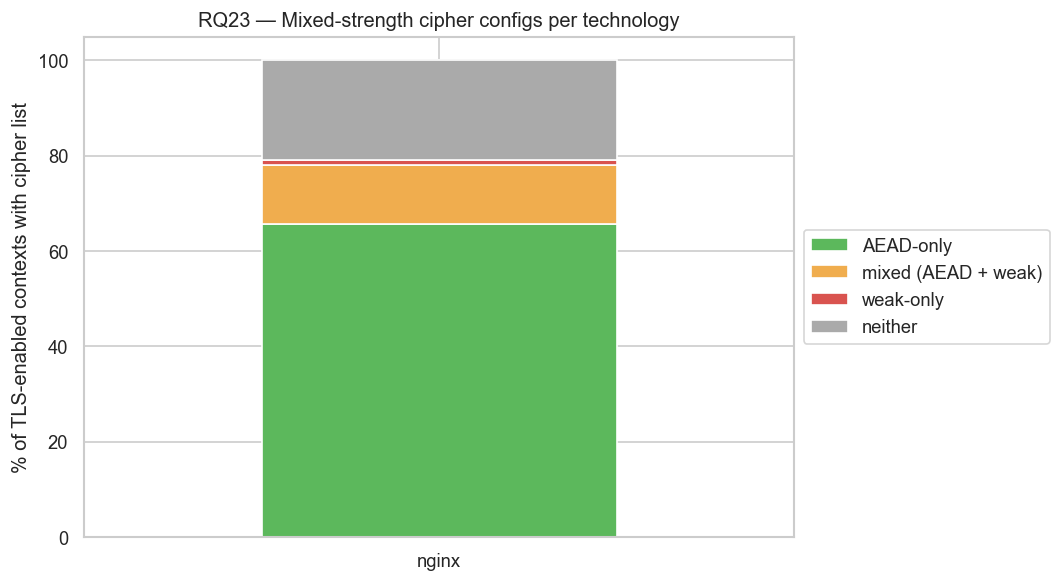}
    \caption{Mixed-Strength Configurations}
    \label{fig:rq23_mixed_strength.png}
\end{figure}

The analysis identifies configs that contain both strong (AEAD) and weak (RC4, DES, 3DES, EXPORT, NULL, MD5) ciphers in the same list — the signature of incremental updates without cleanup. These are arguably more dangerous than purely weak configs because they create a false sense of security. Figure~\ref{fig:rq23_mixed_strength.png}.

\subsubsection{RQ8: Configuration Decay}

\begin{table}[t]
  \centering
  \caption{Comparison of Legacy OpenSSL Presets Across Repository Status}
  \label{tab:legacy_presets_comparison}
  \begin{tabular}{lcc}
    \toprule
    \textbf{Legacy Preset} & \textbf{Active Repos} & \textbf{Archived Repos} \\
    \midrule
    \texttt{!MD5}   & 29.7\% & 41.5\% \\
    \texttt{!aNULL} & 26.7\% & 32.6\% \\
    \texttt{HIGH}   & 23.1\% & 21.0\% \\
    \texttt{!eNULL} & 4.3\%  & 14.3\% \\
    \bottomrule
  \end{tabular}
\end{table}

\begin{figure*}[h]
    \centering
    \includegraphics[width=0.75\linewidth]{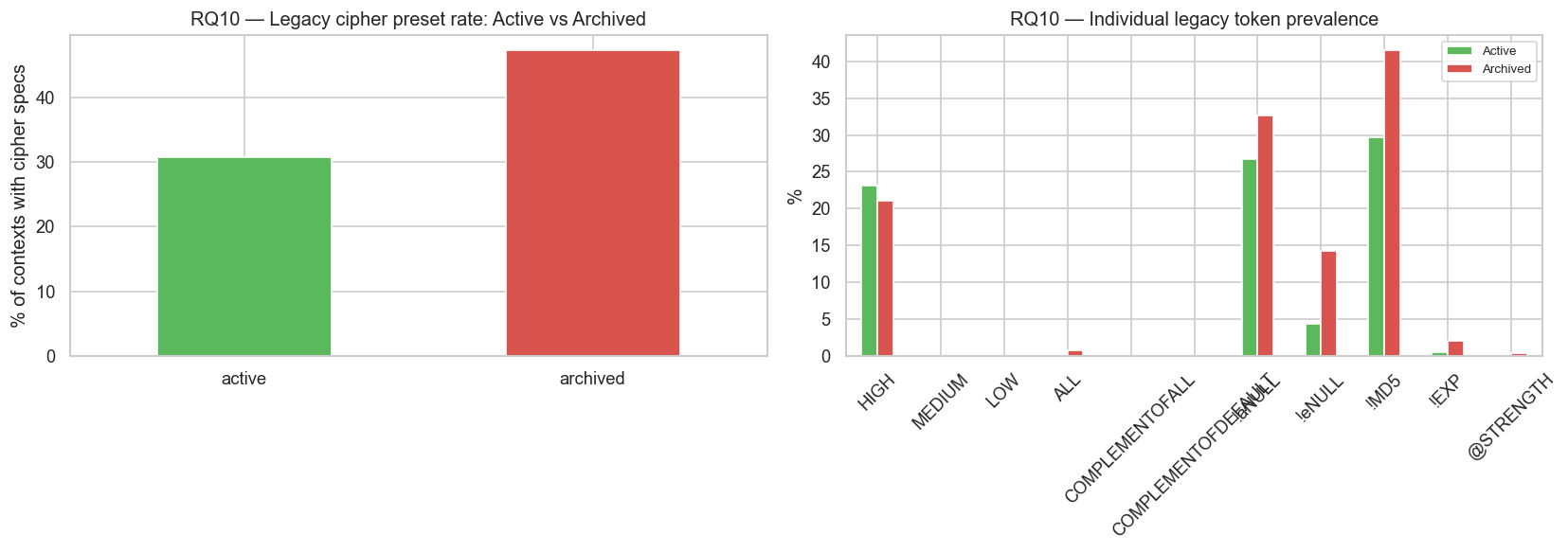}
    \caption{Config Decay}
    \label{fig:rq10_config_decay.png}
\end{figure*}

Before Mozilla's explicit cipher list recommendations became widespread (~2015–2018), the common practice was to use OpenSSL shorthand macros like `HIGH:!aNULL:!MD5:@STRENGTH`. These delegate cipher selection to OpenSSL's version-specific interpretation, meaning the effective cipher list changes silently across OpenSSL upgrades — a maintenance time bomb. Archived repos use `!MD5` and `!aNULL` negation tokens at significantly higher rates — these are hallmarks of the pre-2018 "negate the bad stuff" approach rather than the modern "explicitly list the good stuff" approach. The `!eNULL` gap (4.3\% vs 14.3\%) is particularly stark, showing that active repos have largely moved past this pattern.
 Figure~\ref{fig:rq10_config_decay.png}.

\subsubsection{RQ9: Internal Configuration Consistency}

Many production Nginx configs define multiple `server {}` blocks for different virtual hosts. If these blocks have different TLS settings, it means the operator manages security policy at the vhost level rather than the server level — increasing the chance that one vhost falls behind during hardening updates.

Nearly a third of multi-context config files have inconsistent TLS policies across their server blocks. Cipher drift (19.1\%) is the most common — different vhosts within the same file negotiating different cipher suites. This is a strong indicator of incremental, per-vhost configuration changes without whole-file audits.

\begin{table}[t]
  \centering
  \caption{Drift Rate of Configuration Fields Between Active and Archived Repositories}
  \label{tab:drift_rates}
  \begin{tabular}{lr}
    \toprule
    \textbf{Field} & \textbf{Drift Rate} \\
    \midrule
    Ciphers              & 19.1\% \\
    HSTS                 & 17.4\% \\
    Protocols            & 14.3\% \\
    Prefer server order  & 0.2\%  \\
    \bottomrule
  \end{tabular}
\end{table}

\subsubsection{RQ10: Active vs Archived Drift}

\begin{figure*}[h]
    \centering
    \includegraphics[width=0.75\linewidth]{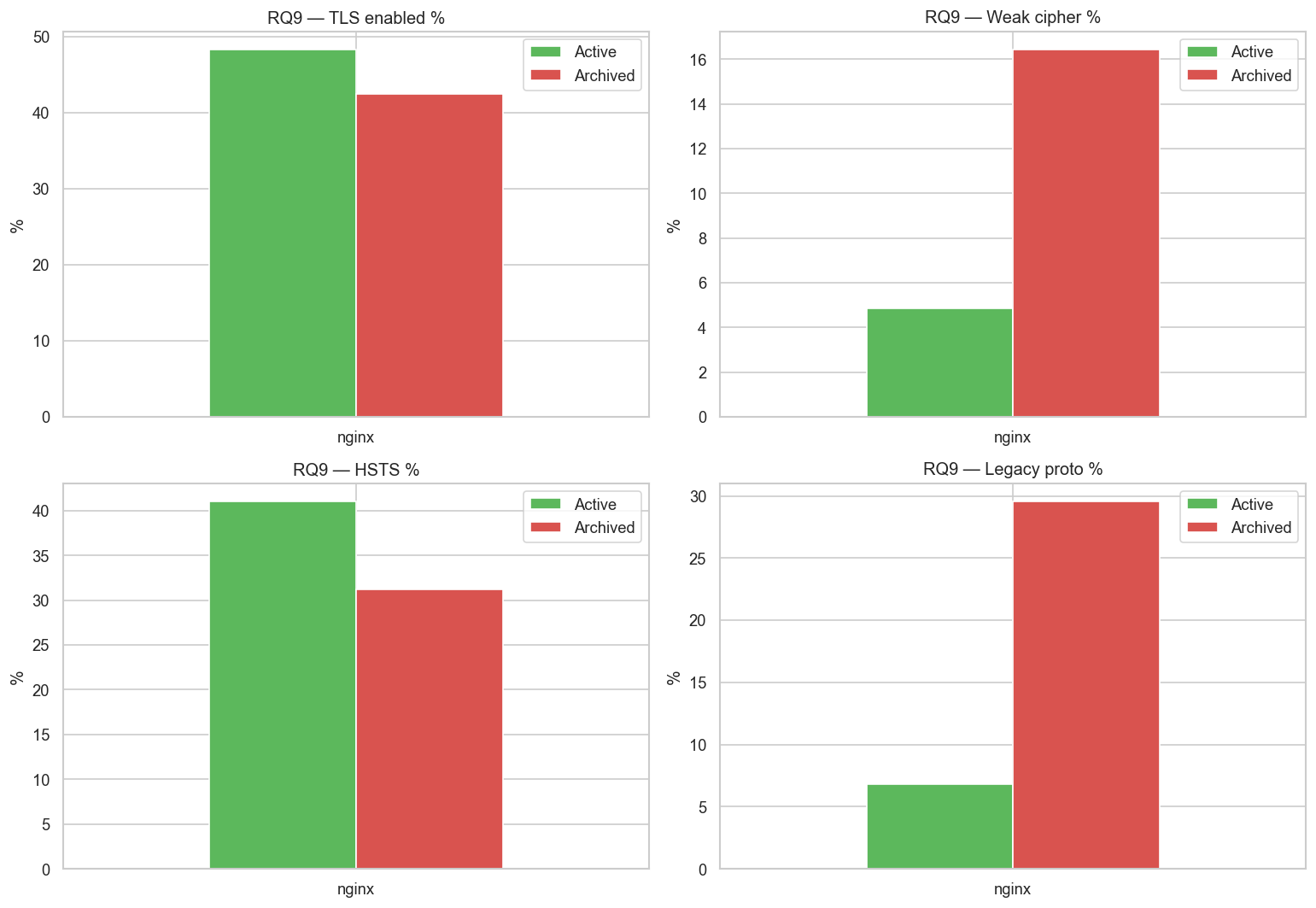}
    \caption{Configuration Decay}
    \label{fig:rq9_active_vs_archived.png}
\end{figure*}

The data confirms that archived repos exhibit worse TLS hygiene across multiple dimensions. Legacy cipher preset usage is notably higher in archived repos (47.3\%) compared to active ones (30.8\%). This 16.5 percentage-point gap demonstrates that TLS configuration quality does improve over time in maintained repositories — but abandoned configs become increasingly dangerous as the cryptographic landscape evolves around them.
Figure~\ref{fig:rq9_active_vs_archived.png}.

\subsubsection{RQ11: AEAD vs CBC Cipher Distribution}

Of configs with explicit cipher lists:

- 72.3\% are AEAD-only — no CBC ciphers present at all
- 6.1\% still include CBC ciphers
- Average AEAD share across all configs: 58.1\%
- Average CBC share: 0.8\%

\begin{figure*}[h]
    \centering
    \includegraphics[width=0.75\linewidth]{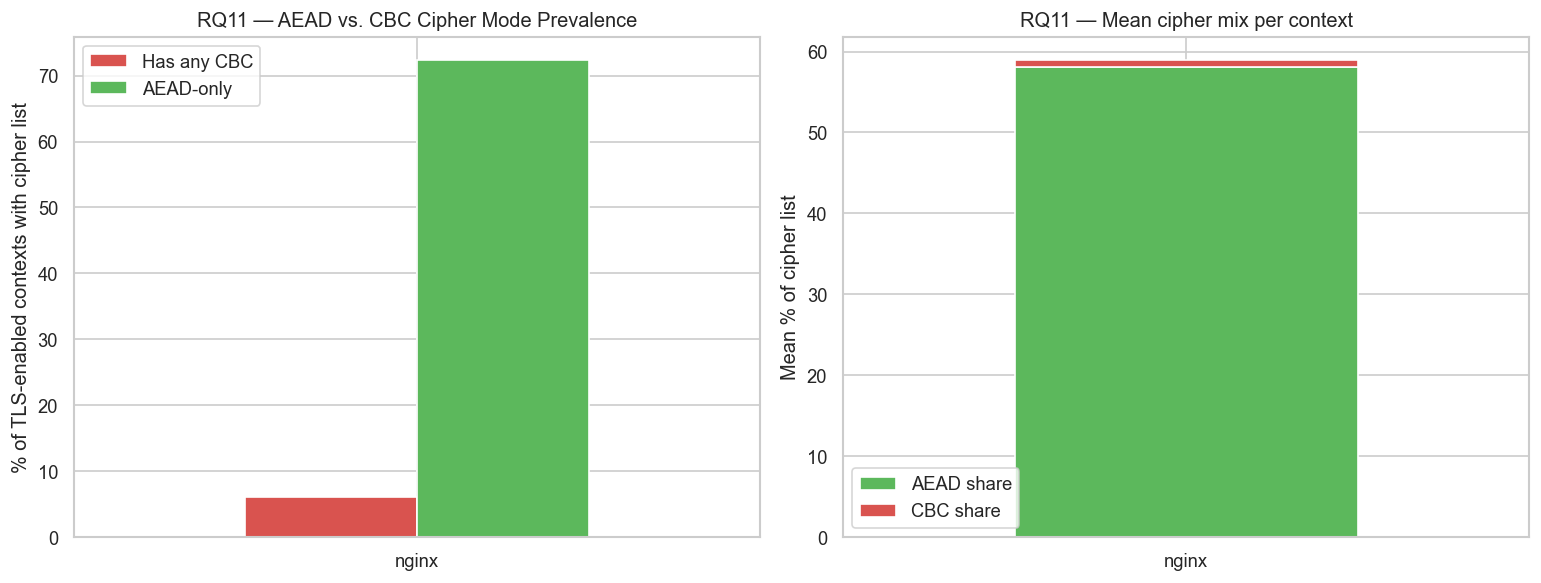}
    \caption{AEAD vs CBC Cipher Distribution}
    \label{fig:rq11_aead_vs_cbc.png}
\end{figure*}

The industry has largely moved to AEAD, but the 6.1\% CBC tail represents configs that haven't been updated since at least 2014 — when POODLE should have prompted immediate CBC removal. Figure~\ref{fig:rq11_aead_vs_cbc.png}.

\subsubsection{RQ12: Cipher Order Hygiene}

\begin{figure*}[h]
    \centering
    \includegraphics[width=0.75\linewidth]{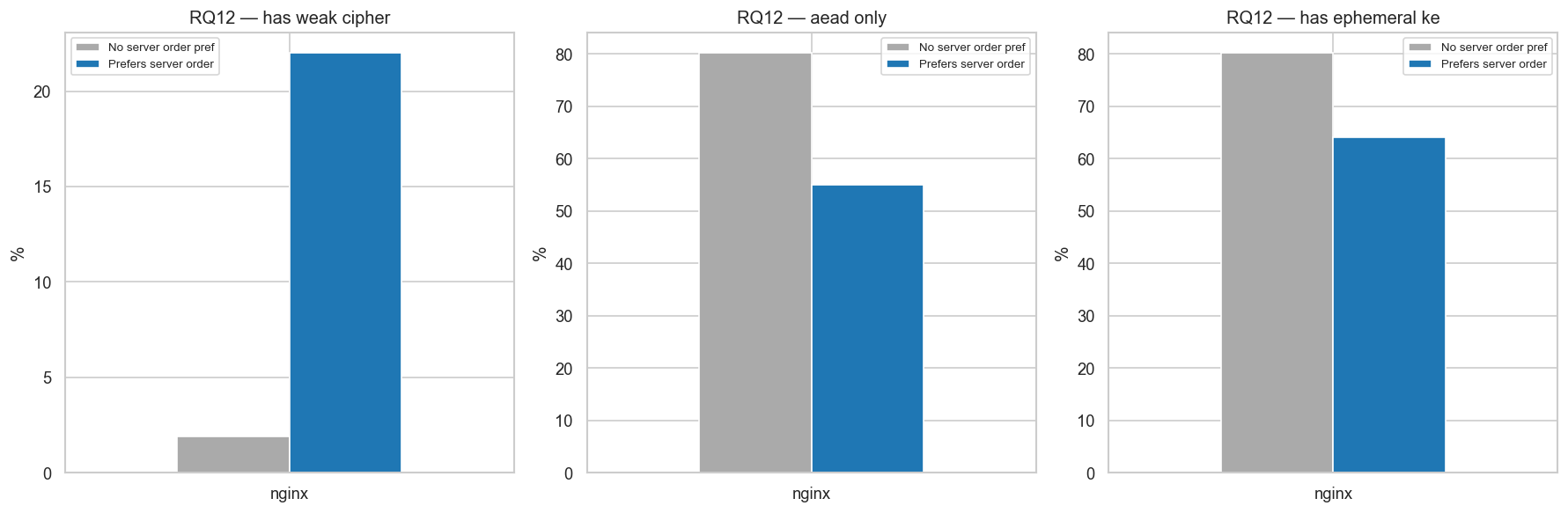}
    \caption{Cipher Order Hygiene}
    \label{fig:rq12_cipher_order_hygiene.png}
\end{figure*}

37.8\% of TLS-enabled contexts set `ssl\_prefer\_server\_ciphers`. The analysis examines whether this directive correlates with actual cipher list quality — whether operators who set it also curate their cipher lists, or whether it's applied mechanically as part of a configuration template without understanding its purpose. Figure~\ref{fig:rq12_cipher_order_hygiene.png}.

\subsubsection{RQ13: ECDHE Named Curve Preferences}

\begin{figure*}[!t]
    \centering
    \includegraphics[width=0.8\linewidth]{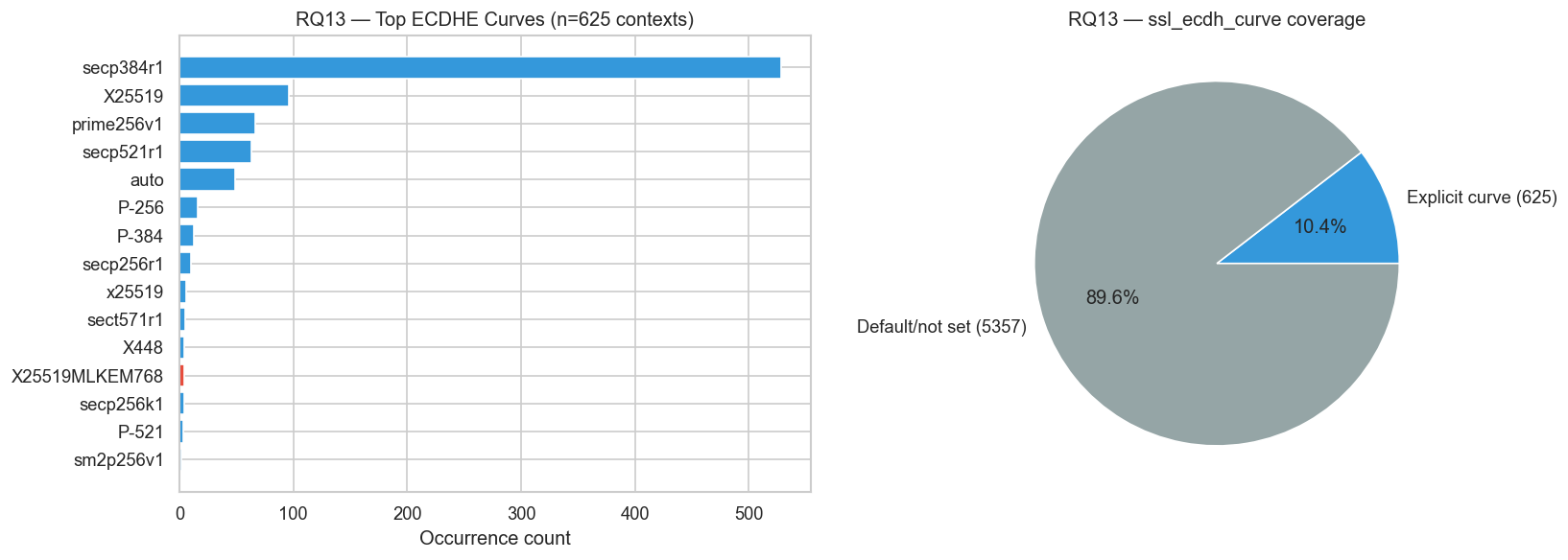}
    \caption{ECDHE Named Curve Preferences}
    \label{fig:rq13_ecdh_curves.png}
\end{figure*}

Only 10.4\% of TLS-enabled contexts (625 of 5,982) explicitly set `ssl\_ecdh\_curve`. Among those:

\begin{table}[t]
  \centering
  \caption{Distribution of Configured Elliptic Curves}
  \label{tab:curve_distribution}
  \begin{tabular}{lr}
    \toprule
    \textbf{Curve} & \textbf{Count} \\
    \midrule
    \texttt{secp384r1}           & 528 \\
    \texttt{X25519}               & 96  \\
    \texttt{prime256v1}           & 67  \\
    \texttt{secp521r1}           & 63  \\
    \texttt{auto}                 & 49  \\
    \texttt{X25519MLKEM768} (PQC) & 4   \\
    \bottomrule
  \end{tabular}
\end{table}

secp384r1 dominates — likely because Mozilla's Intermediate profile recommends it. X25519 is present but uncommon. The PQC hybrid curve `X25519MLKEM768` appears in exactly 4 configs (0.8\% of curve-specifying contexts) — the earliest adopters of post-quantum key exchange in the wild. Figure~\ref{fig:rq13_ecdh_curves.png}.

\subsubsection{RQ14: Cipher String Length}

\begin{figure}[h]
    \centering
    \includegraphics[width=\linewidth]{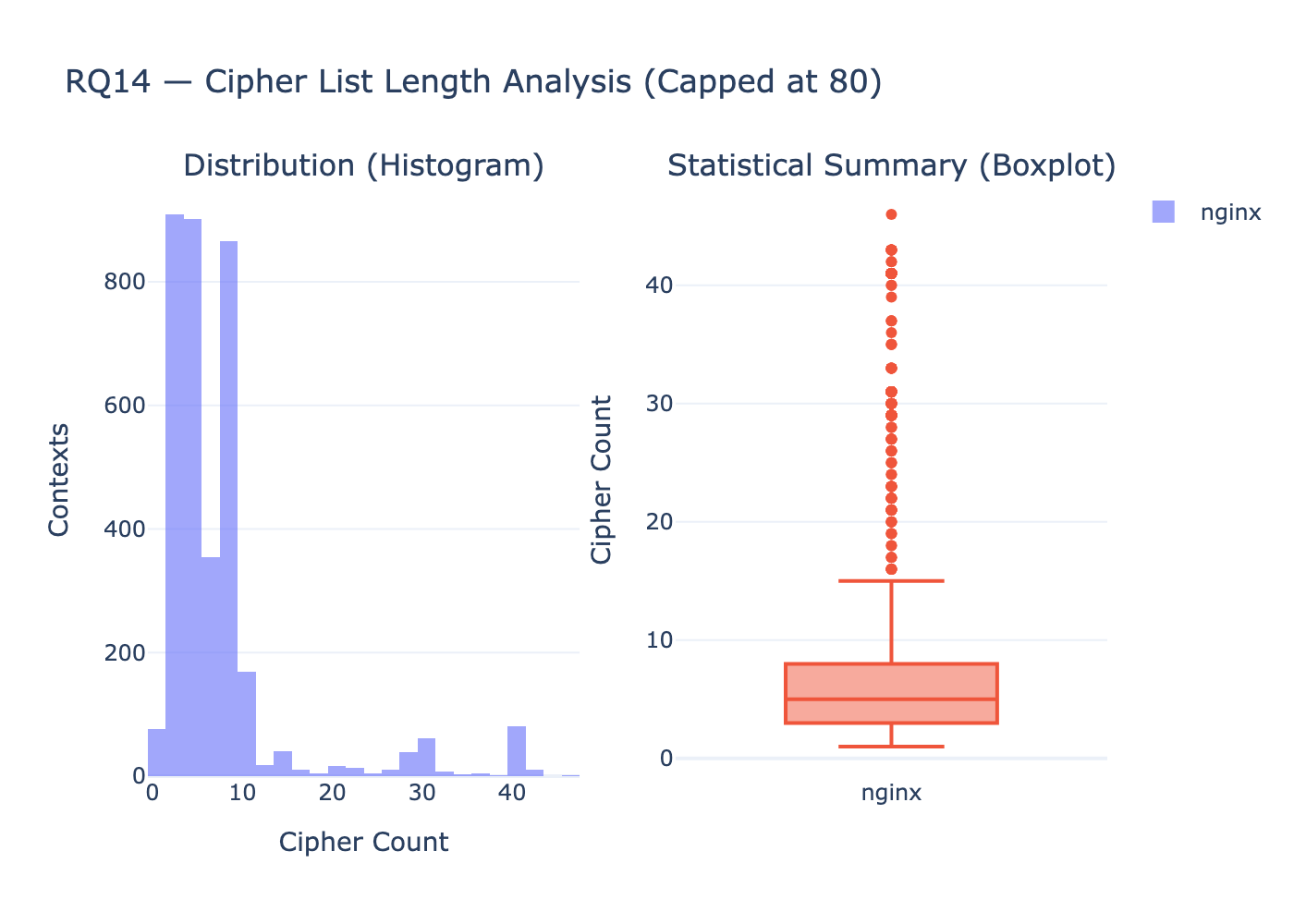}
    \caption{Cipher String Length}
    \label{fig:rq14_cipher_length.png}
\end{figure}

The distribution reveals whether the Nginx community tends toward minimal, security-focused cipher lists or sprawling legacy configurations accumulated over years of incremental changes. Figure~\ref{fig:rq14_cipher_length.png}.

\subsubsection{RQ15: Session Management}

\begin{table}[t]
  \centering
  \caption{Distribution of TLS Session Cache Types}
  \label{tab:session_cache}
  \begin{tabular}{lr}
    \toprule
    \textbf{Cache Type} & \textbf{Share} \\
    \midrule
    Not set (default)   & 50.9\% \\
    Shared              & 47.6\% \\
    Builtin             & 1.3\%  \\
    Off                 & 0.2\%  \\
    \bottomrule
  \end{tabular}
\end{table}

\begin{figure*}[h]
    \centering
    \includegraphics[width=0.75\linewidth]{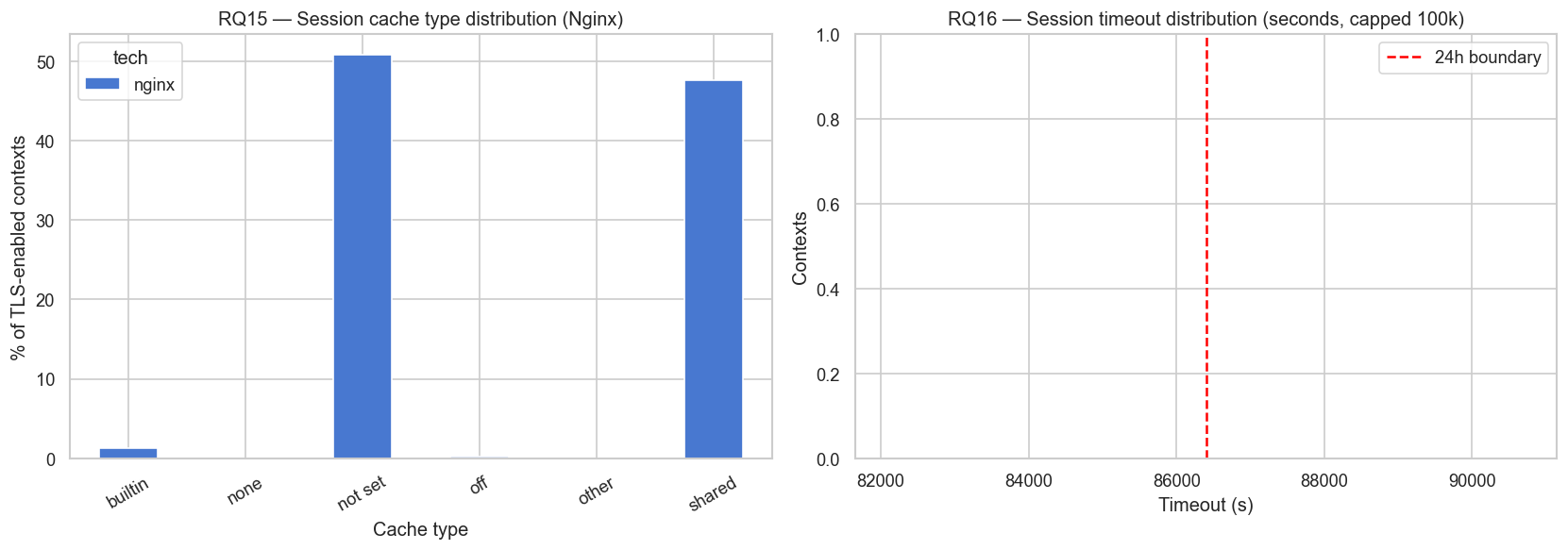}
    \caption{Session Management}
    \label{fig:rq15_rq16_session.png}
\end{figure*}

Half of configs rely on Nginx's default session cache behavior, while 47.6\% explicitly configure shared caches — the production-recommended pattern. Figure~\ref{fig:rq15_rq16_session.png}.

\subsubsection{RQ16: Let's Encrypt Adoption}

\begin{table}[t]
  \centering
  \caption{Distribution of Certificate Origins}
  \label{tab:cert_origin}
  \begin{tabular}{lr}
    \toprule
    \textbf{Certificate Origin} & \textbf{Share} \\
    \midrule
    Other (traditional CA, unclassifiable) & 58.3\% \\
    Let's Encrypt / ACME                   & 34.4\% \\
    Self-signed indicator                  & 4.5\%  \\
    Environment variable                   & 2.9\%  \\
    \bottomrule
  \end{tabular}
\end{table}

Let's Encrypt powers over a third of TLS-enabled Nginx contexts in public repos — a remarkable achievement for a service that didn't exist before 2016. The 4.5\% self-signed rate is relatively low, suggesting that free CA certificates have substantially reduced the incentive to self-sign.

\begin{figure}[h]
    \centering
    \includegraphics[width=\linewidth]{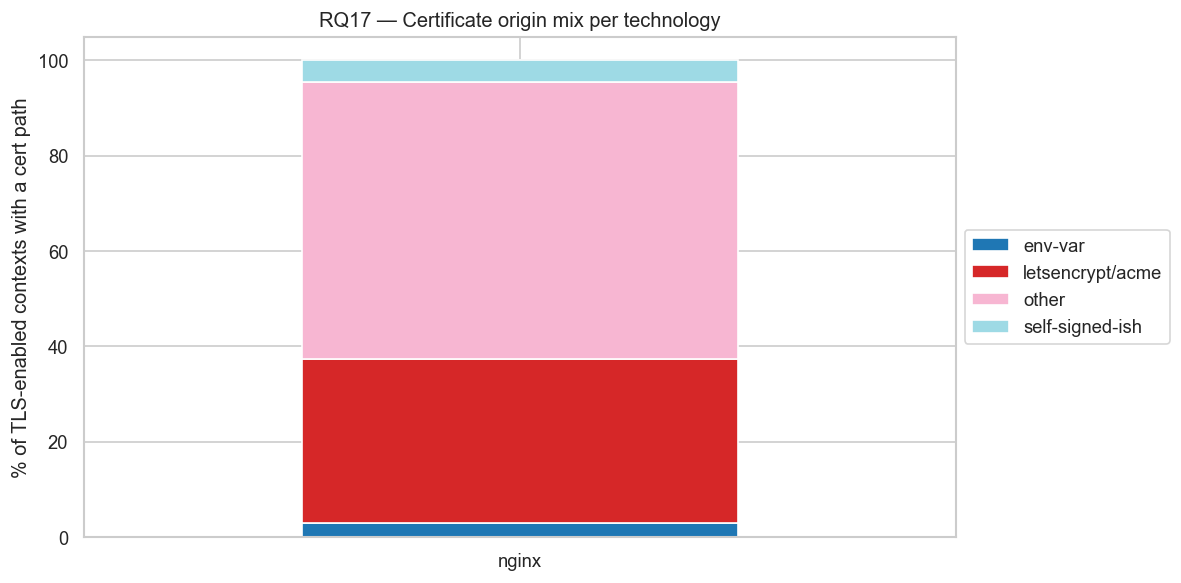}
    \caption{Let's Encrypt Adoption}
    \label{fig:rq17_letsencrypt.png}
\end{figure}

Figure~\ref{fig:rq17_letsencrypt.png}.

\subsubsection{RQ17: Certificate Bundle vs Leaf-Only}

\begin{table}[t]
  \centering
  \caption{Analysis of Certificate Chain Configurations}
  \label{tab:cert_chain_types}
  \begin{tabular}{lrr}
    \toprule
    \textbf{Type} & \textbf{Count} & \textbf{Share} \\
    \midrule
    Leaf-only        & 3,028 & 53.3\% \\
    Fullchain bundle & 2,656 & 46.7\% \\
    \bottomrule
  \end{tabular}
\end{table}

\begin{figure}[h]
    \centering
    \includegraphics[width=\linewidth]{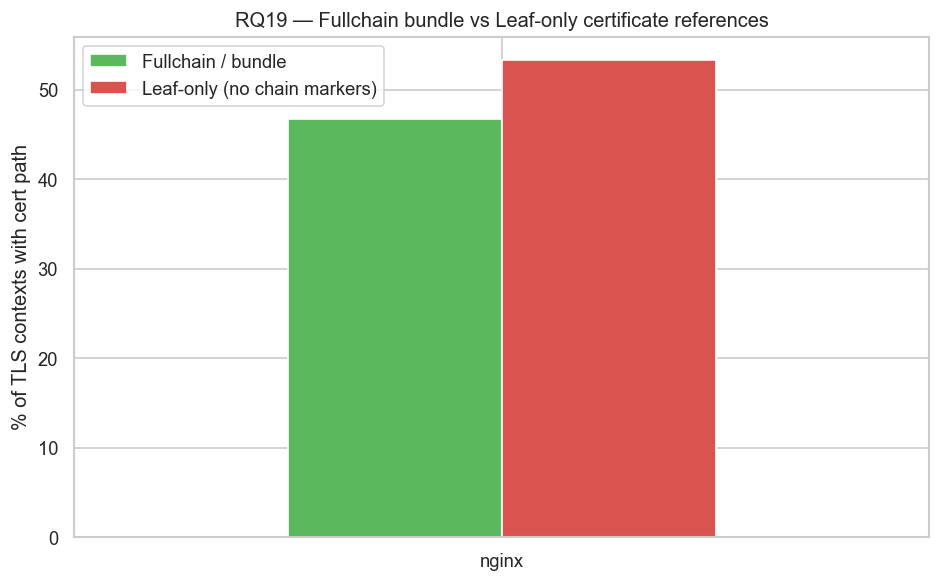}
    \caption{Certificate Bundle vs Leaf-Only}
    \label{fig:rq19_cert_bundle.png}
\end{figure}

More than half of configs (53.3\%) reference only the leaf certificate — meaning they may cause chain validation failures on strict clients. Among fullchain configs, Let's Encrypt paths (`/etc/letsencrypt/live/*/fullchain.pem`) dominate, confirming that ACME tooling encourages correct certificate bundling practices. Figure~\ref{fig:rq19_cert_bundle.png}.

\subsubsection{RQ18: Certificate Storage Patterns}

\begin{table}[t]
  \centering
  \caption{Storage Patterns for Certificate and Key Files}
  \label{tab:storage_patterns}
  \begin{tabular}{lr}
    \toprule
    \textbf{Storage Pattern} & \textbf{Share} \\
    \midrule
    \texttt{/etc/nginx/}               & 34.2\% \\
    \texttt{/etc/letsencrypt/}         & 30.3\% \\
    Other absolute path                & 11.4\% \\
    \texttt{/etc/ssl/} or \texttt{/etc/pki/} & 10.8\% \\
    Environment variable               & 6.9\%  \\
    Relative path                      & 4.9\%  \\
    \texttt{/opt/} or \texttt{/srv/}   & 1.0\%  \\
    Docker secrets                     & 0.5\%  \\
    \bottomrule
  \end{tabular}
\end{table}

\begin{figure}[h]
    \centering
    \includegraphics[width=\linewidth]{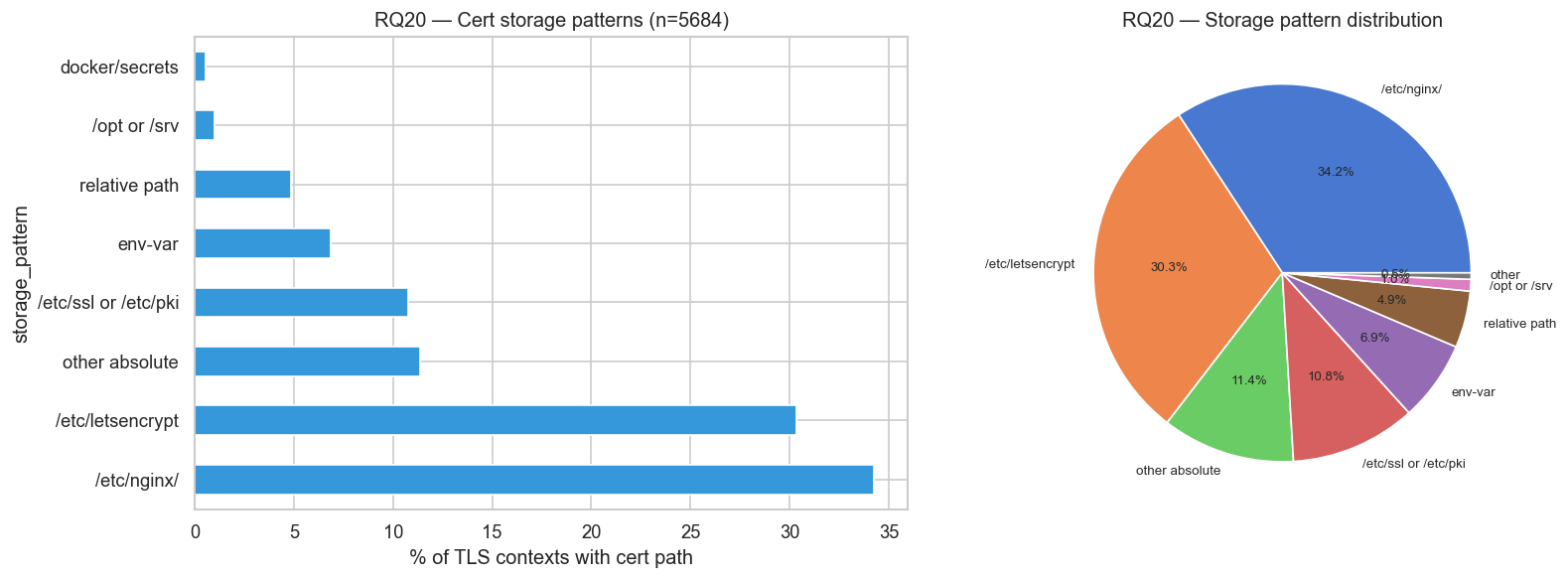}
    \caption{Certificate Storage Patterns}
    \label{fig:rq20_cert_storage.png}
\end{figure}

Nginx's own directory (`/etc/nginx/`) is the most common storage location, followed closely by Let's Encrypt's standard path. The 6.9\% environment-variable rate and 0.5\% Docker secrets rate quantify the container-native deployment segment. Figure~\ref{fig:rq20_cert_storage.png}.

\subsubsection{RQ19: Plaintext Listener Coexistence}

- Files with TLS contexts: 4,755
- Files also containing non-TLS contexts: 3,424 (72.0\%)
- TLS-only files: 1,331

\begin{figure}[h]
    \centering
    \includegraphics[width=\linewidth]{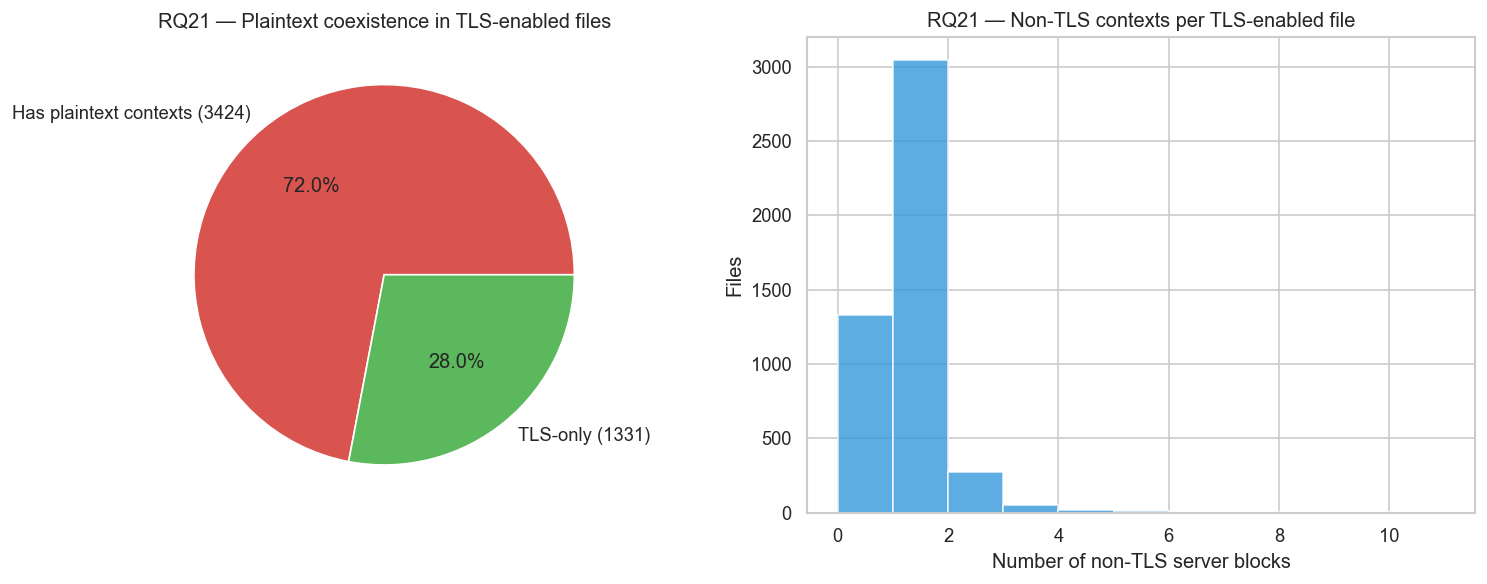}
    \caption{Plaintext Listener Coexistence}
    \label{fig:rq21_plaintext_coexistence.png}
\end{figure}

Nearly three-quarters of TLS-enabled config files also define plaintext listeners. While many of these port 80 blocks likely redirect to HTTPS (a best practice), the prevalence highlights that pure TLS-only configurations are the minority. Without examining each port 80 block for redirect logic, we cannot distinguish intentional redirect blocks from genuine plaintext content serving. Figure~\ref{fig:rq21_plaintext_coexistence.png}.

\subsubsection{RQ20: Default and Wildcard Server Names}

\begin{table}[t]
  \centering
  \caption{Analysis of Configured Nginx Hostnames (\texttt{server\_name})}
  \label{tab:hostname_types}
  \begin{tabular}{lr}
    \toprule
    \textbf{Hostname Type} & \textbf{Share} \\
    \midrule
    Domain-like (production)         & 51.4\% \\
    \texttt{localhost}               & 10.3\% \\
    Empty / not set                  & 10.3\% \\
    \texttt{\_} (catch-all)          & 9.4\%  \\
    Environment variable (\texttt{\$\{...\}}) & 8.3\%  \\
    Other                            & 5.5\%  \\
    \texttt{example.com}             & 3.8\%  \\
    Wildcard (\texttt{*})            & 0.5\%  \\
    \texttt{127.0.0.1}               & 0.5\%  \\
    \bottomrule
  \end{tabular}
\end{table}

\begin{figure}[h]
    \centering
    \includegraphics[width=\linewidth]{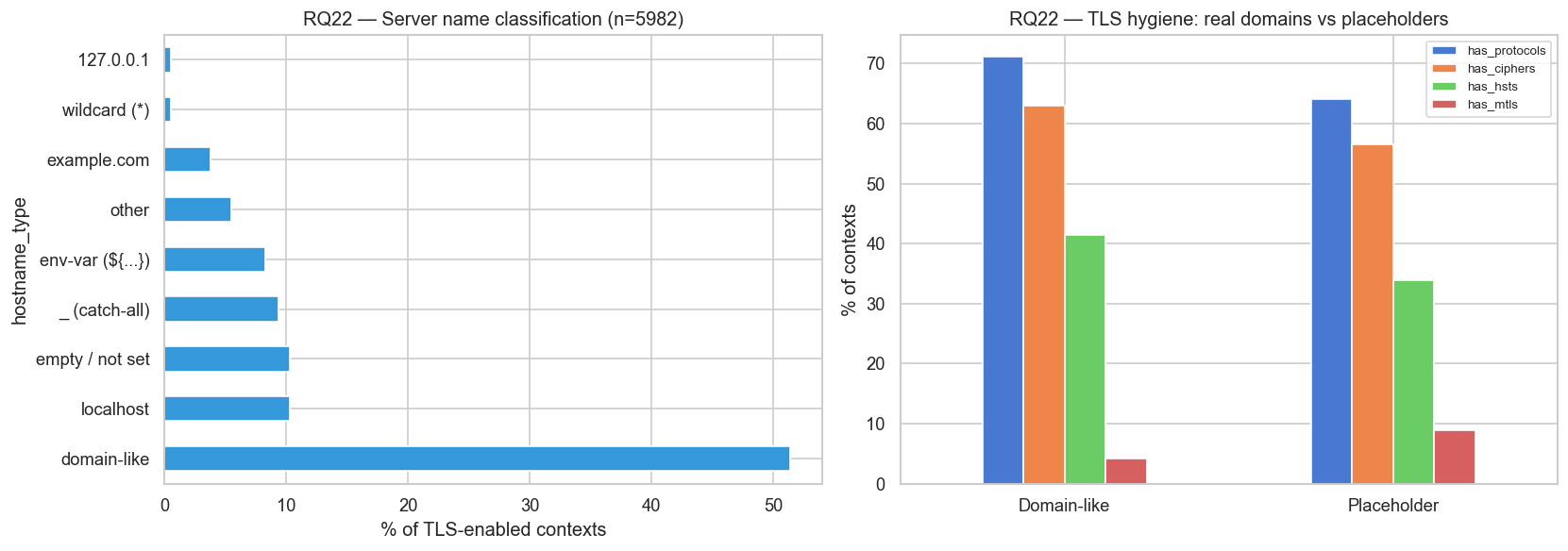}
    \caption{Default and Wildcard Server Names}
    \label{fig:rq22_server_names.png}
\end{figure}

42.6\% of TLS-enabled contexts use likely non-production hostnames. This is a significant finding for interpreting the entire dataset: nearly half the configs are templates, tutorials, dev environments, or placeholder configurations — not hardened production systems. The 8.3\% environment-variable rate represents configs designed for deployment-time hostname injection (Docker, Kubernetes). Figure~\ref{fig:rq22_server_names.png}.

\subsection{Key Takeaways}

Two headline observations follow directly from this study. First, the quantum vulnerability surface is universal: every measured configuration relies on quantum-vulnerable key exchange, and 28.9\% additionally lack forward secrecy. Second, the config-level view reveals what scanners miss: 21.8\% of configurations permit TLS~1.0/1.1 despite likely negotiating only TLS~1.2/1.3 under normal
conditions.

The finding that 28.9\% of real-world TLS configurations in the wild rely on
RSA key exchange with no forward secrecy implies that in an organisation's context, any session recorded by a network-position adversary can be retroactively decrypted without requiring a quantum computer at all, compromise of the server's long-term private key is sufficient. Migration to ephemeral key
exchange should therefore be treated as an immediate action item independent of quantum timelines, with hybrid PQC key exchange
as the natural next step.

The finding that 29.7\% of configurations delegate TLS settings
entirely to framework defaults has direct implications for
deployment standards. Organisations that embed hybrid
PQC cipher suites into approved baseline templates for Nginx,
Apache, and Java services can achieve broad migration coverage
through template governance rather than per-system
remediation, consistent with the structural observation that
template adoption drives posture at scale.

\subsection{Limitations}

\paragraph{Dataset representativeness.}

The measurement corpus consists of public GitHub repositories,
which overrepresent development and template configurations:
42.6\% of TLS-enabled contexts use non-production hostnames.
Production financial services configurations are not publicly
available.

\paragraph{Technology coverage.}

The framework covers Nginx, Apache, and Spring Boot. HAProxy,
Envoy, IIS, Tomcat, and cloud-managed TLS services are not
represented. Extending coverage to these technologies is a
direct avenue for future work.

\paragraph{Key exchange only.}
Both the measurement study and the deployment focus on TLS key
exchange. Certificate migration (RSA/ECDSA to ML-DSA) is not
addressed and warrants separate study once IETF LAMPS composite
certificate standards stabilize.

\paragraph{Single institution deployment.}

The OCBC case study represents a single institution's deployment.
Generalizability to other institutions, architectures, or
higher-traffic environments should be validated through additional
deployments.

\section{Hybrid PQC Deployment in a Production Banking Environment}
\label{sec:casestudy}

\subsection{Deployment Context and Threat Model}

We follow up with our proposed approach of TLS discovery with practical PQC deployment in internal services gateways like web servers and load balancers
in a PoC banking environment. The deployment targets an internal banking application recording sensitive financial transactions. 
It is a system whose data carries long-term confidentiality requirements, making it a priority target for HNDL-style quantum attacks.

The threat model has two components. The \textit{future} threat is a
CRQC running Shor's algorithm~\cite{shor1994}, capable of breaking
the ECDH key exchange in the system's existing TLS connections. The
\textit{present} threat is an adversary archiving encrypted TLS
sessions today with the intent to decrypt them retroactively once
quantum capability matures~\cite{tno2024handbook}. The migration
objective was to protect the key exchange layer of all TLS sessions
touching the internal banking application, while preserving full backward
compatibility with existing clients and making zero changes to the
banking application itself.

\subsection{System Architecture}

\begin{figure}[h]
    \centering
    \includegraphics[width=\linewidth]{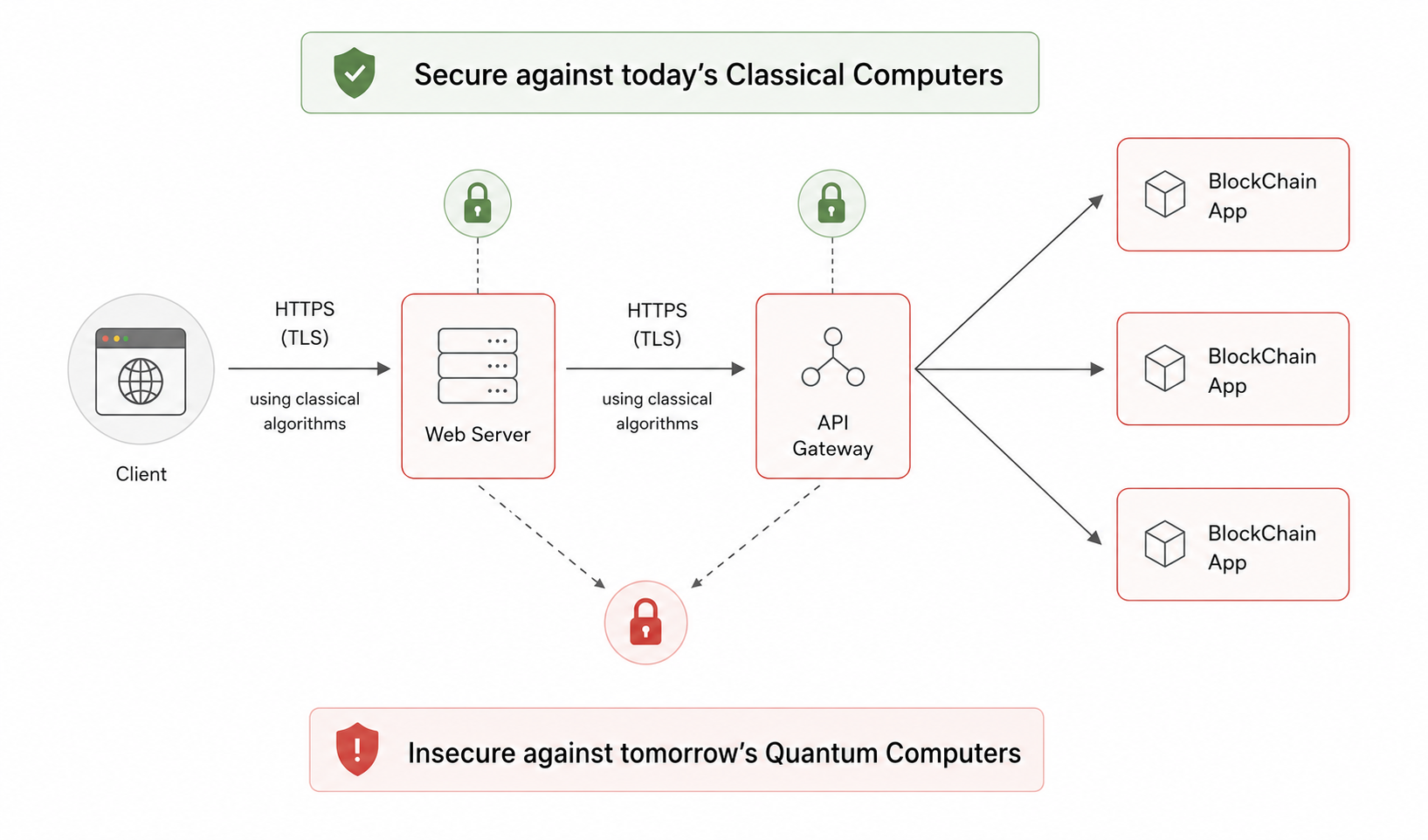}
    \caption{The Three Tier Architecture}
    \label{fig:deployed_architecture}
\end{figure}

The deployment comprises three tiers, as illustrated in
Figure~\ref{fig:deployed_architecture}.

\begin{enumerate}
  \item \textbf{Public-facing web server (Apache).} Terminates TLS
    connections from internet-facing clients; acts as a reverse proxy
    routing requests to the API gateway. Primary TLS termination
    point for external traffic and first migration target.

  \item \textbf{API gateway (Java / Spring Boot).} Manages
    authentication, authorization, rate limiting, and routing.
    Terminates TLS from the web server and establishes new TLS
    sessions toward backend services. Second migration target.

  \item \textbf{Internal Banking application.} Hosts sensitive financial records, and interacts with the API gateway.
\end{enumerate}

The existing TLS relied on ECDHE with P-256 for key exchange, directly
vulnerable to Shor's algorithm. Digital certificate authentication
was retained unchanged in this phase; certificate migration to ML-DSA
is identified as a subsequent phase (Section~\ref{sec:discussion}).

\subsection{Implementation}

\subsubsection{Hybrid PQC at the Web Server Layer}

Introducing ML-KEM support at the Apache layer required upgrading
OpenSSL to incorporate the Open Quantum Safe (OQS) provider~\cite{oqs_provider}, which implements NIST PQC algorithms
as OpenSSL-compatible primitives. The server was then configured to offer hybrid key exchange in TLS~1.3.

The \texttt{X25519MLKEM768} hybrid group combines classical X25519 with ML-KEM-768, providing simultaneous security against classical
and quantum adversaries. Clients that do not support hybrid groups fall back to \texttt{X25519}, ensuring full backward compatibility.

\subsubsection{Hybrid PQC at the API Gateway Layer}

The API gateway operates within the Java ecosystem, where TLS is managed through the Java Cryptography Architecture (JCA) and Java
Secure Socket Extension (JSSE) frameworks. By default, these frameworks support only classical algorithms. To introduce ML-KEM
support, the BouncyCastle PQC security provider~\cite{bouncycastle} was registered as a JCA provider at JVM startup:

\begin{lstlisting}[language=Java,
  caption={BouncyCastle PQC provider registration (abbreviated).},
  label={lst:java}]
Security.insertProviderAt(
    new BouncyCastleJsseProvider(), 1);
Security.insertProviderAt(
    new BouncyCastlePQCProvider(), 2);
\end{lstlisting}

With the provider registered, TLS configuration specifying \texttt{X25519MLKEM768} or \texttt{MLKEM512} key exchange groups is handled 
transparently by the JCA/JSSE layer, with no modification to the gateway application code.

\subsubsection{Deployment and Containerization}

All components were containerized using Docker to ensure configuration consistency across environments and to support
repeatable, scalable deployment. Each container image packages the upgraded OpenSSL or JVM environment alongside the application,
eliminating environment-specific dependency mismatches.

\subsection{Performance Evaluation}

\subsubsection{Methodology}

\begin{figure*}[!t]
    \centering
    \includegraphics[width=0.4\linewidth]{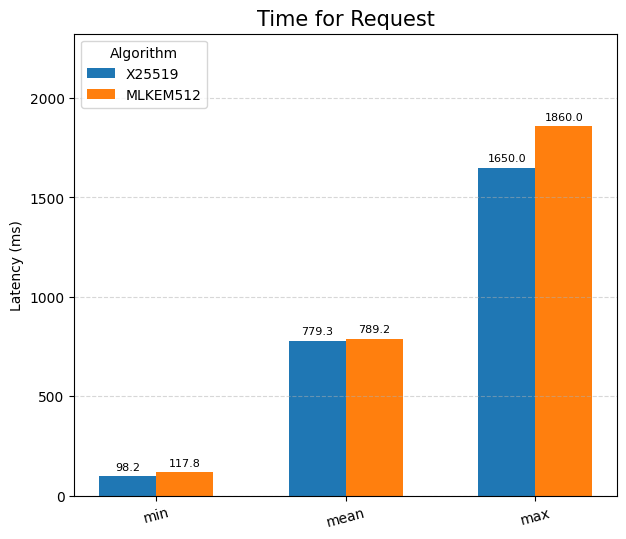}
    \caption{Time For Request}
    \label{fig:tFR.png}
\end{figure*}

\begin{figure*}[!t]
    \centering
    \includegraphics[width=0.4\linewidth]{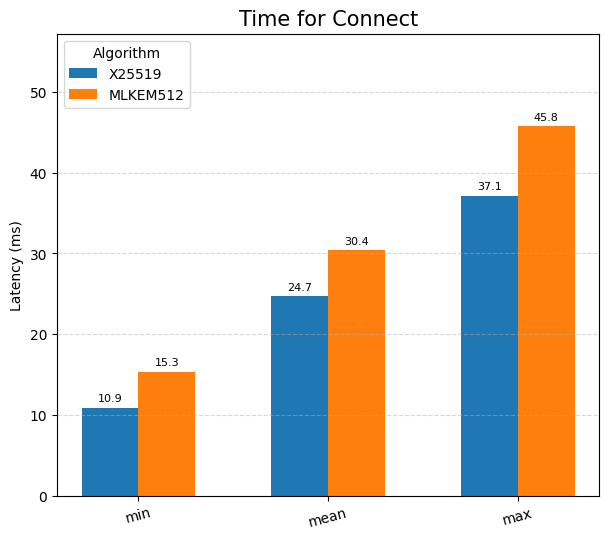}
    \caption{Time to Connect}
    \label{fig:tFC.png}
\end{figure*}

\begin{figure*}[!t]
    \centering
    \includegraphics[width=0.4\linewidth]{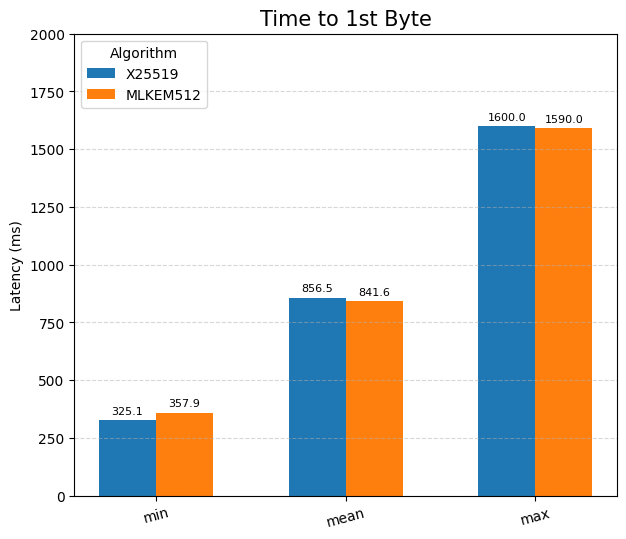}
    \caption{Time to 1st Byte}
    \label{fig:tF1stByte.png}
\end{figure*}

We evaluated performance impact using \texttt{h2load}, the HTTP/2 load testing tool from the nghttp2 project, with $C = 100$
concurrent clients over a 30-second duration. We compared two configurations: (1) classical X25519 as the baseline, and (2)
ML-KEM-512 hybrid key exchange as the PQC-enabled configuration. Four metrics were measured: time-for-request (end-to-end latency) 
Figure~\ref{fig:tFR.png}, time-for-connect (TCP + TLS handshake time) Figure~\ref{fig:tFC.png}, time-to-first-byte (server 
processing latency) Figure~\ref{fig:tF1stByte.png}, and per-client request rate.

\subsubsection{Results}

\begin{table}[t]
  \centering
  \caption{h2load Performance: X25519 vs.\ ML-KEM-512
    ($C=100$, 30\,s)}
  \label{tab:perf}
  \setlength{\tabcolsep}{4pt}
  \begin{tabular}{llcccc}
    \toprule
    \textbf{Config} & \textbf{Metric} &
      \textbf{Min} & \textbf{Max} &
      \textbf{Mean} & \textbf{SD} \\
    \midrule
    \multirow{4}{*}{X25519}
      & Time-for-request (ms)   & 98.2  & 1650 & 779.3 & 160.8 \\
      & Time-for-connect (ms)   & 10.9  & 37.2 & 24.7  & 8.2   \\
      & Time-to-1st-byte (ms)   & 325.1 & 1600 & 856.5 & 296.0 \\
      & Reqs/client/s           & 1.12  & 1.40 & 1.27  & 0.06  \\
    \midrule
    \multirow{4}{*}{ML-KEM-512}
      & Time-for-request (ms)   & 117.8 & 1860 & 789.2 & 182.5 \\
      & Time-for-connect (ms)   & 15.3  & 45.8 & 30.4  & 10.0  \\
      & Time-to-1st-byte (ms)   & 357.9 & 1590 & 841.6 & 276.1 \\
      & Reqs/client/s           & 1.10  & 1.37 & 1.25  & 0.06  \\
    \bottomrule
  \end{tabular}
\end{table}

The primary overhead is in connection establishment: mean time-for-connect increases from 24.7\,ms to 30.4\,ms
(approximately +23\%), reflecting the additional computational
work of ML-KEM key encapsulation during the TLS handshake.
End-to-end request latency shows a modest mean increase of
approximately 1.3\%. Per-client throughput decreases by
approximately 1.6\%. The deployment produced \textbf{zero
errors} across all test runs under both configurations. This
overhead is an acceptable and expected trade-off for quantum-safe
security in a financial institution context where connection reuse
is standard and protected data carries decade-long confidentiality
requirements.

\subsection{Lessons Learned}

\paragraph{Layer-by-layer migration is practical.}

Securing the web server and API gateway layers required no changes
to the internal banking application. In enterprises where applications
are accessed through API gateways or load balancers, migrating
the termination layer first maximizes coverage with minimal
application-level disruption.

\paragraph{Provider-level integration enables Java ecosystem migration.}

The JCA/JSSE provider architecture allows PQC support to be
injected into Java environments without modifying application
code. BouncyCastle's PQC provider~\cite{bouncycastle} makes this
injection straightforward, but operators must explicitly register
it at JVM startup; it is not enabled by default.

\paragraph{HSM readiness must be assessed early.}

This deployment used filesystem-backed private keys. In production
environments where keys are HSM-backed, the HSM's firmware must
be assessed for ML-KEM support before migration can proceed. HSM
readiness assessment should be the first action for any financial
institution beginning a PQC migration program.

\paragraph{Certificate migration remains an open phase.}

This deployment addresses key exchange only; certificates continue
to use RSA or ECDSA signatures. Migration to ML-DSA certificates
is the following necessary phase, which will serve as future work, once there is 
support of PQC signatures in the supporting PKI.

\subsection{Discussion: Synthesizing the Three Contributions}
\label{sec:discussion}

The three contributions address different phases of the same
underlying problem. The discovery framework (C1) provides
the tooling to answer the same question for a specific
organization's own infrastructure. The measurement study (C2) presents a large-scale empirical measurement of 8,443 real‑world Nginx TLS configurations from public repositories, establishing a baseline of TLS hygiene and quantum vulnerability surface across protocol, cipher suite, key exchange, and certificate management practices in the wild. The OCBC deployment (C3) demonstrates what executing the migration looks like in practice.
Together, they form a coherent arc from tooling to measurement to execution.

A unifying observation is that the primary barrier to quantum-safe
migration is not algorithmic readiness but operational visibility.
NIST has published final standards. Reference implementations exist.
Hybrid cipher suites are available in current versions of OpenSSL
and BouncyCastle~\cite{oqs_provider,bouncycastle}. What
organizations lack is a precise, machine-readable answer to what
they are currently running, and therefore what they need to change, and how to effect that change, at scale.

\section{Conclusion and Future Work}
\label{sec:conclusion}

The post-quantum migration of enterprise TLS infrastructure is not a
future problem, it is a present one. Reference PQC implementations
are available, however the main barrier is operational: organizations do
not know, with precision, what cryptography their infrastructure is
currently configured to use, and without that knowledge, no migration
program can be planned or executed with confidence.

This paper has addressed that barrier through three complementary
contributions. We presented an automated, deterministic framework for
extracting and normalizing TLS cryptographic configuration directly
from Nginx, Apache, and Spring Boot infrastructure, producing a unified,
provenance-traced inventory suitable for quantum-risk assessment and
policy comparison. We reported a large-scale empirical measurement
study of 8,443 real-world Nginx configurations, establishing that
post-quantum hybrid key exchange adoption is zero in the measured
corpus, that 28.9\% of configurations rely on RSA key exchange with
no forward secrecy, and that 21.8\% still permit deprecated
TLS~1.0 or~1.1, a baseline against which future adoption can be
measured. And we documented a practical hybrid PQC
deployment in modern service gateways like web servers and API gateways, demonstrating that
ML-KEM-512 and X25519-MLKEM768 can be introduced at the TLS termination layer of a multi-tier banking architecture with manageable performance overhead, zero application-layer changes, 
and zero errors under load.

The central lesson across all three contributions is architectural:
TLS termination is distributed across a cryptographic perimeter of
load balancers, API gateways, reverse proxies, and web servers, each of which is an independent migration target. We have demonstrated that we can achieve config-level visibility, and a migration strategy that begins at the outermost termination
layer and works inward. The framework and deployment described in this paper provide both.

We consider the following tasks as future work: Extending config-level TLS discovery to HAProxy, Envoy, and cloud-native TLS services would substantially increase coverage of the enterprise cryptographic perimeter. A longitudinal measurement study tracking PQC adoption in public repositories over time would provide a quantitative baseline for monitoring 
industry migration progress; the four \texttt{X25519MLKEM768} configurations identified in this corpus represent the leading edge of what will become a measurable trend. 
The certificate migration phase, and deploying ML-DSA certificates alongside hybrid key exchange, warrants a dedicated study once standards stabilize and we have support for 
PQC signautures in PKI.

\section*{Acknowledgment}

This work was supported by Oversea-Chinese Banking Corporation Limited
(OCBC Bank), Singapore. The authors thank Praveen Raina, Peter Koh,
Mayda Lim, and Alena Lim for their in-kind contributions and support
throughout the project. The authors also acknowledge the support of
Nanyang Technological University, Singapore.

This work was also supported by the CyberSG R\&D Programme Office, Singapore. The authors thank the programme office team for their guidance and support throughout the project. 
The authors also acknowledge the support of Digital Trust Centre and Nanyang 
Technological University, Singapore.

\bibliographystyle{IEEEtran}
\bibliography{refs}

\balance

\end{document}